\documentclass[11pt]{article}
\usepackage{amsmath}
\usepackage{amssymb}
\usepackage{amsfonts}
\usepackage{epsfig}
\usepackage{calc}
\usepackage{subfigure}

\usepackage[usenames]{color}
\usepackage[normalem]{ulem}

\long\def\comment#1{}

\newfont{\bbb}{msbm10 scaled 700}

\newfont{\bb}{msbm10 scaled 1100}
\newcommand{\CC}{\mbox{\bb C}}

\newcommand{\EE}{\mbox{\bb E}}

\newcommand{\cv}{{\bf c}}

\newcommand{\ev}{{\bf e}}

\newcommand{\hv}{{\bf h}}

\newcommand{\pv}{{\bf p}}

\newcommand{\uv}{{\bf u}}
\newcommand{\wv}{{\bf w}}

\newcommand{\xv}{{\bf x}}
\newcommand{\yv}{{\bf y}}

\newcommand{\zerov}{{\bf 0}}

\newcommand{\Am}{{\bf A}}
\newcommand{\Bm}{{\bf B}}

\newcommand{\Dm}{{\bf D}}

\newcommand{\Gm}{{\bf G}}
\newcommand{\Hm}{{\bf H}}
\newcommand{\Id}{{\bf I}}

\newcommand{\Sm}{{\bf S}}

\newcommand{\Um}{{\bf U}}

\newcommand{\Xm}{{\bf X}}
\newcommand{\Ym}{{\bf Y}}

\newcommand{\Ac}{{\cal A}}

\newcommand{\Cc}{{\cal C}}

\newcommand{\Ec}{{\cal E}}

\newcommand{\Kc}{{\cal K}}

\newcommand{\Nc}{{\cal N}}

\newcommand{\Xc}{{\cal X}}

\newcommand{\alphav}{\hbox{\boldmath$\alpha$}}

\newcommand{\lambdav}{\hbox{\boldmath$\lambda$}}

\newcommand{\Lambdam}{\hbox{\boldmath$\Lambda$}}

\newcommand{\Sigmam}{\hbox{\boldmath$\Sigma$}}

\renewcommand{\det}{{\hbox{det}}}

\renewcommand{\arg}{{\hbox{arg}}}

\newcommand{\mmse}{{\sf mmse}}
\newcommand{\zf}{{\sf zf}}

\newcommand{\eqdef}{\stackrel{\Delta}{=}}

\newcommand{\herm}{{\sf H}}

\newtheorem{thm}{Theorem}

\newcommand{\beq}{\begin{equation}}
\newcommand{\eeq}{\end{equation}}

\newcommand{\bea}{\begin{eqnarray}}
\newcommand{\eea}{\end{eqnarray}}
\newcommand{\bean}{\begin{eqnarray*}}
\newcommand{\eean}{\end{eqnarray*}}

\newcommand{\bit}{\begin{itemize}}
\newcommand{\eit}{\end{itemize}}
\newcommand{\ben}{\begin{enumerate}}
\newcommand{\een}{\end{enumerate}}

\title{Asymptotic Performance of Linear Receivers in MIMO Fading Channels}
\author{K. Raj Kumar$^\dagger$, G. Caire$^\dagger$ and A. L. Moustakas$^\star$
\thanks{The material in this paper was presented in part at the IEEE Information
Theory Workshop (ITW-07), Lake Tahoe, USA, Sep.~2-6, 2007. This
research was supported in part by the European Commission under
Grant "PHYSCOM" with No. MIRG-CT-2005-030833 and by an Oakley
fellowship from the University of Southern California.}}
\date{\today}

\def\BibTeX{{\rm B\kern-.05em{\sc i\kern-.025em b}\kern-.08em
    T\kern-.1667em\lower.7ex\hbox{E}\kern-.125emX}}

\begin{document}
\maketitle

\vskip 4cm \noindent $\dagger$
Department of EE - Systems, University of Southern California\\
Los Angeles, CA 90007, USA\\
E-mail: {\tt rkkrishn@usc.edu}, {\tt caire@usc.edu} \\[12pt]
$\star$ Department of Physics, National \& Capodistrian Univ. of Athens\\
Athens, Greece\\
E-mail: {\tt arislm@phys.uoa.gr}

\vskip 3cm \noindent {\bf Keywords:}
Diversity Multiplexing Tradeoff, Large-System Limit, Linear Receivers, MIMO Channels, Spatial Multiplexing.

\clearpage
\newpage

\begin{abstract}
Linear receivers are an attractive low-complexity alternative to
optimal processing for multi-antenna MIMO communications. In this
paper we characterize the information-theoretic performance of MIMO
linear receivers in two different asymptotic regimes. For fixed
number of antennas, we investigate the limit of error probability in
the high-SNR regime in terms of the Diversity-Multiplexing Tradeoff
(DMT). Following this, we characterize the error probability for
fixed SNR in the regime of large (but finite) number of antennas.

As far as the DMT is concerned, we report a negative result: we show that both linear Zero-Forcing (ZF)
and linear Minimum Mean-Square Error (MMSE) receivers achieve the same DMT, which is
largely suboptimal even in the case where outer coding and decoding is performed across the antennas.
We also provide an approximate quantitative analysis of the markedly
different behavior of the MMSE and ZF receivers at finite rate and
non-asymptotic SNR, and show that while the ZF receiver achieves poor diversity at any
finite rate, the MMSE receiver error curve slope flattens out progressively,
as the coding rate increases.

When SNR is fixed and the number of antennas becomes large,
we show that the mutual information at the output of a MMSE or ZF linear receiver
has fluctuations that converge in distribution to a Gaussian random variable,
whose mean and variance can be  characterized in closed form.
This analysis extends to the linear receiver case a well-known
result previously obtained for the optimal receiver.
Simulations reveal that the asymptotic analysis captures accurately
the outage behavior of systems even with a moderate number of antennas.
\end{abstract}

\clearpage
\newpage

\section{Introduction}
\label{sec:intro}

The next generation of wireless communication systems is expected to
capitalize on the large gains in spectral efficiency and reliability
promised by MIMO multi-antenna communications
\cite{Tel,ZheTse,book1,book2} and include MIMO technology as a
fundamental component of their physical layer \cite{ieee802.11n}.
The information theoretic analysis and the efficient design of
space-time (ST) codes for transmission over these MIMO systems have
been active areas of research over the past decade. Also, suboptimal
low-complexity receiver schemes have been widely proposed and
investigated as a low-complexity alternative to the optimal
Maximum-Likelihood (ML) or ML-like receivers
\cite{DamGamCai,MurGamDamCai}. These schemes range from the
iterative interference (soft) cancellation (e.g.,
\cite{hesham-iterative}), to successive interference (hard)
cancellation (e.g., \cite{vblast,wstc}), to the even lower
complexity ``separated'' architecture, based on linear spatial
equalization followed by standard single-input single-output (SISO)
decoding.\footnote{It should be noticed that the current MIMO WLAN
standard \cite{ieee802.11n} is based on MIMO-OFDM, therefore, linear
equalization is performed in the space and in the frequency domains.
For simplicity, in this work we restrict ourselves to the standard
frequency-flat case where equalization is purely spatial.}

In this paper, we present two types of asymptotic performance
analysis of this low-complexity MIMO architecture. First, we consider
the Diversity-Multiplexing Tradeoff (DMT) \cite{ZheTse},
which captures the performance tradeoff between rate and
block-error probability in the high-SNR, high spectral efficiency regime.
We determine the DMT achieved by low-complexity MIMO architectures that use Zero-Forcing (ZF) or
Minimum Mean-Square Error (MMSE) linear receivers and apply
conventional SISO outer coding {\em before} the MIMO transmitter
and conventional SISO decoding to the output of the linear
receiver. The DMT analysis reveals that both ZF and MMSE linear
receivers are very suboptimal in terms of their achievable diversity.
Furthermore, we observe that while the DMT analysis accurately
predicts the behavior of the ZF receiver at all finite rates, the
performance of the MMSE receiver is in stark contrast to that
predicted by the DMT analysis at low rates.
In fact, we observe that for sufficiently low rates the MMSE receiver exhibits an ML-like performance.
On the contrary, when working at higher rates (and correspondingly higher SNR) the MMSE receiver
approaches the ZF performance. We provide an approximate analysis that explains this behavior both
qualitatively and quantitatively.

In the second part of this paper we take a closer look at the
performance of the linear MMSE and ZF receivers at finite SNR.
Since this is very difficult to capture in closed form, we explore
a second type of asymptotic regime, where we fix SNR and let the
number of antennas become large. Using random matrix theory, we
show that in this case the limiting distribution of the mutual
information of the parallel channels induced by the linear receiver
is Gaussian, with mean and variance that can be computed in closed form.
The analysis provides accurate results even for a moderate number of antennas
and allows to quantify how the performance loss in terms of diversity suffered by linear receivers
may be recovered by increasing the number of antennas.
This prompts to the conclusion that  in order to achieve
a desired target spectral efficiency and block-error rate,
at given SNR and receiver complexity, increasing the number of antennas and using simple
linear receiver processing may be, in fact, a good design option.

The paper is organized as follows. In the rest of this section we
briefly comment on concurrent existing literature.
In Section \ref{system-model}, we define the system model and recall the main
facts the ZF and MMSE linear receivers considered in this work.
Section \ref{DMT-analysis} presents the DMT analysis and some illustrative numerical examples.
Section \ref{sec:MMSE_finite_rate} is devoted to the fixed-rate analysis
of the MMSE receiver performance with coding across the antennas
and provides an approximate quantitative analysis of the slope of
the error probability versus SNR.
Section \ref{sec:MMSE_MI_distrib} deals with the
limiting distribution of the mutual information for the MMSE
and ZF receivers for a large number of antennas and provides
some illustrative numerical examples on the validity and limitations of this analysis.
Conclusions are pointed out in Section \ref{conclusions} and some technical
details of the proofs are deferred to the Appendix.

\subsection{Related literature}

Since its introduction in the seminal work \cite{ZheTse}, the DMT
has become a {\em standard tool} in the characterization of the
performance of slowly-varying fading channels in the high-SNR, large
spectral efficiency regime. Space-time coding schemes have been
characterized in terms of their achievable DMT in a series of works,
including lattice coding and decoding \cite{GamCaiDam_LaST}  and ZF
or MMSE {\em decision feedback} receivers (see for example
\cite{JiaVar,JiaVarLi}). The multipath diversity achievable by {\em
linear equalizers} in frequency-selective SISO channels has also
attracted some attention and was recently solved in
\cite{Nos-isit07}. The spatial diversity achievable by MIMO linear
receivers and separated detection and decoding was investigated in
parallel and independently in \cite{itw07} by the authors
\footnote{The present paper provides the detailed proofs of the DMT
results presented in \cite{itw07} and presents the novel
large-system finite-SNR analysis of the MMSE receiver, which is not
given in \cite{itw07}.} and in \cite{HedNos}. In this respect, it is
worthwhile to stress the differences between the present work and
\cite{HedNos}: 1) we investigate the full DMT curve, while
\cite{HedNos} focuses only on the fixed-rate case (corresponding to
zero multiplexing gain); 2) \cite{HedNos} develops only lower bounds
to the diversity order, based on upper bounds on the outage
probability, while we have both lower and upper bounds and show that
they are tight; 3) the analysis on the diversity order of the ZF
receiver in \cite{HedNos} is fundamentally flawed for the case of
coding across the antennas. In fact, \cite{HedNos} conjectures that
the channel gains in the parallel channels induced by the ZF
receiver are {\em statistically independent}. If this was the case,
the diversity order would be very different, as detailed in a comment
at the end of Section \ref{discussion}. Indeed, the final result in \cite{HedNos} is
correct because of a compensation of errors.
In contrast, we show that the channel gains
are strongly correlated, and this is precisely why coding across the
antennas does not buy any extra diversity with respect to pure
spatial multiplexing;  4) in \cite{HedNos} the diversity of the MMSE
receiver with coding across the antennas is characterized in the
region of low rates and high rates for the case of two transmit
antennas. In contrast, the approximate analysis presented in Section
\ref{sec:MMSE_finite_rate} of this paper characterizes the diversity
of the MMSE receiver for the {\em whole range of intermediate rates}
from ``low'' to ``high'' and for arbitrary number of antennas.

With respect to the large-system analysis of linear receivers
presented in Section \ref{sec:MMSE_MI_distrib}, we notice that
asymptotic Gaussianity was shown for the MIMO channel mutual
information given by the ``log-det'' formula, whose cumulative
distribution function (cdf) yields the block-error rate achievable
under {\em optimal} decoding. This was shown in various works, such
as \cite{Hochwald2002_MultiAntennaChannelHardening,
Moustakas2003_MIMO1,
Smith2002_OnTheGaussianApproximationToTheCapacityOfWirelessMIMOSystems,
Hachem2007_NewApproachGaussianMIMO}. At the same time the {\em
marginal} asymptotic Gaussianity of the SINR of a single MMSE and ZF
receiver channel was derived in
\cite{Tse2000_MMSEFluctuations,Liang2007_MMSEAsymptotics}, without
looking at the {\em joint} Gaussianity of all SINRs for all these
channels. While the marginal Gaussianity is useful in the case of
pure spatial multiplexing, where each antenna (or ``spatial
stream'') is independently encoded and decoded, we would like to
remark here that the joint Gaussianity is crucial in the analysis of
the most relevant case where outer coding is applied across
antennas. In Section \ref{sec:MMSE_MI_distrib} we characterize the
limiting joint Gaussian distribution of the SINRs and obtain the
statistics of the mutual information of linear MMSE and ZF receivers
for the case of coding across the antennas. Our approach is novel
and does not follow as a simple extension of the analysis of the
marginal statistics as done previously.

\section{System model,  DMT and linear receivers} \label{system-model}

\begin{figure}[htbp]
\begin{center}
\includegraphics[width=15cm]{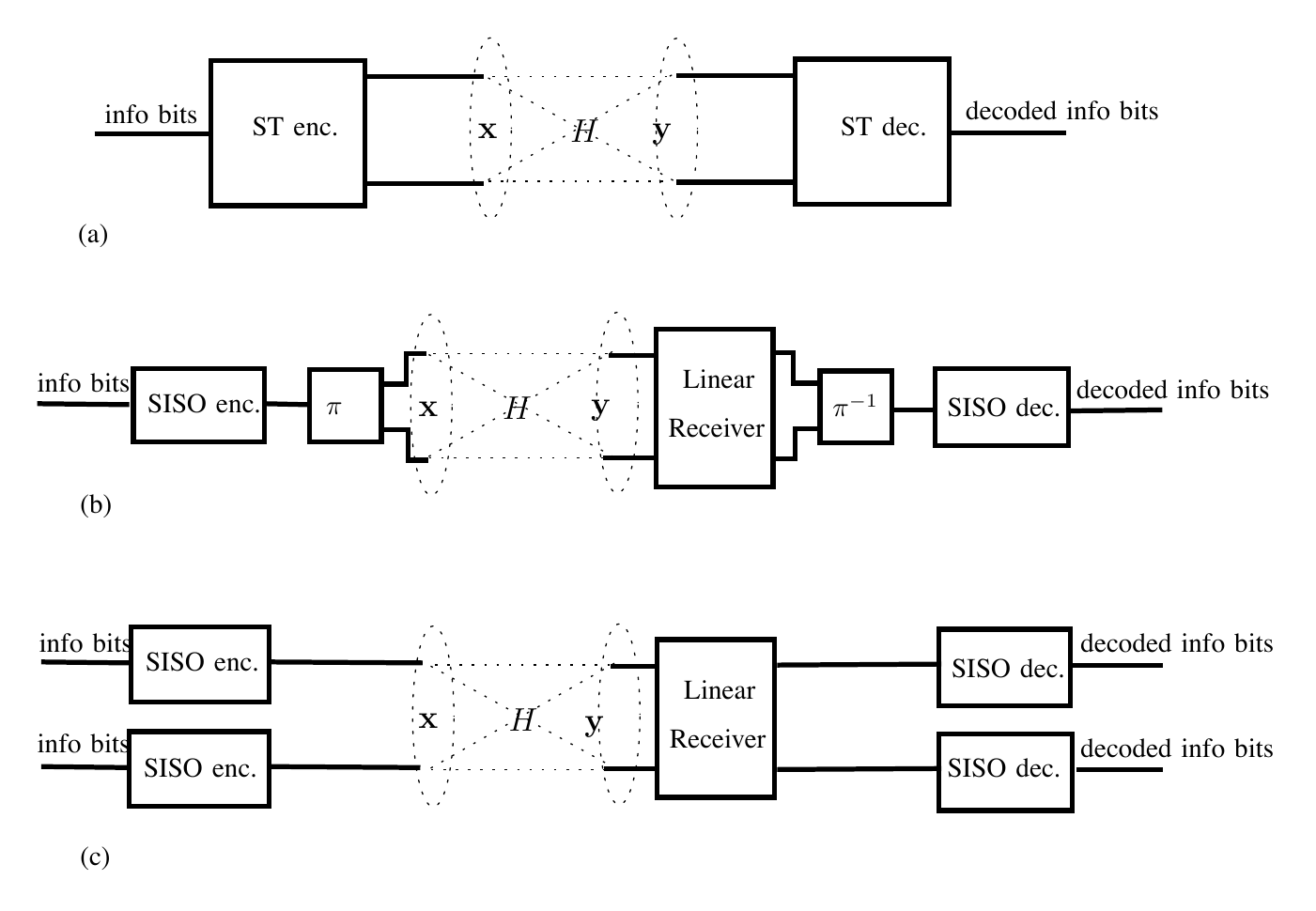}
\caption{Three possible space-time architectures: (a) unrestricted space-time coding scheme; (b) coding across the antennas, with
linear spatial equalization; (c) pure spatial multiplexing with linear spatial equalization.
$\pi$ and $\pi^{-1}$ in (b) denote interleaving and de-interleaving.}
\label{block-diag}
\end{center}
\end{figure}

Fig.~\ref{block-diag} shows three types of MIMO architectures,
employing $M$ transmit and $N$ receive antennas.
Since the focus of this paper is on linear receivers, we shall assume $N \geq M$ throughout this paper.
Scheme (a) puts no restriction on the choice of the space-time coding and decoding
scheme: the $M$ channel inputs are jointly encoded, and the $N$
channel outputs are jointly and possibly optimally decoded. Scheme
(b) is based on interleaving and demultiplexing over the $M$
inputs the codewords of a SISO code. A linear spatial equalizer
(referred briefly as ``linear receiver'' in the following)
processes each $N$-dimensional channel output vector (purely
spatial processing) and creates $M$ virtual {\em approximately
parallel} channels (details are given later on). The output of
these virtual channels are then demultiplexed and deinterleaved,
and eventually fed to a SISO decoder that treats them as scalar
observations, thus disregarding the possible dependencies
introduced by the underlying MIMO channel.
Notice that in scheme (b) coding is applied {\em across the antennas}.
Finally, scheme (c) is based solely on ``spatial multiplexing'', that is, $M$ independently
encoded streams drive the $M$ transmit antennas and are
approximately separated by the linear receiver, the outputs of
which are fed to $M$ independent decoders.

The output of the underlying frequency-flat slowly-varying
MIMO channel is given by
\begin{equation} \label{eq:mimo}
\yv_t = \Hm \xv_t + \wv_t, \;\;\; t = 1,\ldots,T,
\end{equation}
where $\xv_t \in \CC^M$ denotes the channel input vector at
channel use $t$, $\wv_t \sim \Cc\Nc(\zerov, N_0 \Id)$ is the
additive spatially and temporally white Gaussian noise and $\Hm
\in \CC^{N \times M}$ is the channel matrix. In this work we make
the standard assumption that the entries of $\Hm$ are i.i.d. $\sim
\Cc\Nc(0,1)$, and that $\Hm$ is random but constant over the
duration $T$ of a codeword (quasi-static Rayleigh i.i.d. fading
\cite{Tel,ZheTse}). The input is subject to the total power constraint
\begin{equation} \label{short-term-power-constraint}
\frac{1}{MT} \EE \left[ \|\Xm\|_F^2 \right] \leq E_s,
\end{equation}
where $\Xm = [\xv_1,\ldots,\xv_T]$ denotes a space-time codeword,
uniformly distributed over the space-time codebook $\Xc$, and
$\|\cdot\|_F$ denotes the Frobenius norm. Furthermore, following the
standard literature of MIMO channels and space-time coding, we
define the transmit SNR $\rho$ as the total transmit energy per
time-slot over the noise power spectral density, i.e., $\rho = M
E_s/N_0$.

We assume no Channel State Information (CSI) at the transmitter.
In this work we consider the case of very large block length (and
consequently of very slowly-varying fading).  Under the
quasi-static assumption, it is well-known that the capacity and
the outage capacity (or $\epsilon$-capacity) are independent of
the assumption on CSI at the receiver \cite{bps}. Hence, assuming
perfect CSI at the receiver incurs no loss of generality.

We focus on the MIMO detector/decoder blocks in
Fig.~\ref{block-diag}. Under the fully unconstrained ST
architecture (a), the optimum receiver for the MIMO channel in
\eqref{eq:mimo} is the maximum likelihood (ML) decoder, with
minimum distance decision rule given by
\[ \hat{\Xm} = \arg\min\limits_{\Xm \in \Xc} \| \Ym - \Hm \Xm \|^2_F. \]
This entails joint processing of the symbols across all antennas at
the receiver, over the whole block length $T$, and is typically
implemented using algorithms like Sphere Decoding (see
\cite{DamGamCai} and reference therein) and their tree search
sequential decoding generalization \cite{MurGamDamCai}, possibly
coupled with ML Viterbi algorithm if the underlying code has a
trellis structure (e.g., \cite{KumCai_slast,viterbo-tcm}). The
performance of this decoder is characterized by the information outage
probability given by
\begin{equation} \label{eq:outage}
P_{\rm out}(R,\rho) = \inf\limits_{
\begin{array}{c}
\Sm :\Sm \succeq 0\\
\text{tr}(\Sm)\leq 1
\end{array}} P\left ( \log \det (I + \rho \Hm \Sm \Hm^\herm ) \leq R \right ).
\end{equation}
where the optimization is over the Hermitian symmetric non-negative definite matrix
$\Sm$ subject to a trace constraint, reflecting the channel input power constraint
(\ref{short-term-power-constraint}).
Several lower complexity suboptimal decoders have been proposed in the literature.
In particular, architectures (b) and (c) in Fig.~\ref{block-diag}
involve a linear {\em memoryless} receiver defined by the matrix
$\Gm$, such that the output of the linear receiver is $\yv'_t = \Gm
\yv_t$. Classical choices for $\Gm$ are the ZF or the MMSE spatial
filters, or any diagonal scaling thereof. Under the assumption of
Gaussian inputs, very large block length $T$ and ideal interleaving,
the linear receiver creates $M$ ``virtual'' parallel channels that,
without loss of generality, can be described by
\begin{equation} \label{parallel-ch}
y'_{k,t} = \sqrt{\gamma_k} x_{k,t}  + w'_{k,t}, \;\; k =
1,\ldots,M,
\end{equation}
where we normalize the input and output such that $\EE[|x_{k,t}|^2]
= \EE[|w'_{k,t}|^2] = 1$, and where $\gamma_k$ denotes the Signal to
Interference plus Noise Ratio (SINR) at the $k$-th linear receiver
output.\footnote{In order to avoid any misunderstanding, it should
be noticed here  that ``interference'' is uniquely  caused by the
generally non-perfect separation of the transmitted symbols in
$\xv_t$ by the linear receiver $\Gm$.  We consider a strictly
single-user setting, with no multiuser interference.}

Under the above assumptions, the performance of such schemes is
characterized by the following two outage probabilities. With
coding across antennas (scheme (b)), the outage probability of interest is given by
\begin{equation} \label{eq:pout_lin}
P_{\rm out}^{\text{lin}}(R,\rho) \triangleq P \left (\sum_{k=1}^{M} \log (1
+ \gamma_k) \leq R \right );
\end{equation}
Under pure spatial multiplexing (scheme (c)), the relevant outage
probability is given by
\begin{equation} \label{eq:pout_sp_mult}
P_{\rm out}^{\text{sp mult}}(R,\rho) \triangleq P \left ( \bigcup_{k=1}^{M}
\left \{  \log (1 + \gamma_k) \leq \frac{R}{M} \right \} \right ).
\end{equation}
where we used the fact that, by symmetry,
without CSI at the transmitter the optimal performance of spatial
multiplexing with linear receivers is achieved by allocating the
same rate $R/M$ to each stream.

For completeness and for later use, we recall here the expressions
of the SINRs for the ZF and the MMSE linear receivers.

\paragraph{ZF receiver.}
In this case, the matrix $\Gm$ is chosen as $\Gm = \Dm \Hm^+$,
where $\Dm$ is a suitable diagonal scaling matrix and $\Hm^+$ is
the Moore-Penrose pseudo-inverse of $\Hm$ \cite{Ver}. Since $\Hm$
has rank $M$ with probability 1, this takes on the form
\[ \Hm^+ = (\Hm^\herm \Hm)^{-1} \Hm^\herm. \]
In the absence of transmitter CSI, the signal power is allocated
uniformly across the transmitter antennas. It is immediate to show
that the SINRs on the resulting $M$ parallel channels are given by
\begin{equation} \label{sinr-zf}
\gamma_k = \frac{\rho/M}{[(\Hm^\herm \Hm)^{-1}]_{kk}},
\end{equation}
where the notation $[\Am]_{kk}$ indicates the $k^{\text{th}}$
diagonal entry of a matrix $\Am$.

\paragraph{MMSE receiver.}
In this case, the matrix $\Gm$ is chosen in order to maximize the
SINR $\gamma_k$ for each $k$, over all linear receivers. It is
well-known that this is achieved by choosing $\Gm = \Dm \Hm_{\rm
mmse}$, where $\Dm$ is a suitable diagonal scaling matrix and
$\Hm_{\rm mmse}$ is the linear MMSE filter \cite{Ver} that
minimizes the MSE $\EE[\|\xv_t - \Hm_{\rm mmse} \yv_t\|^2]$. Using
the orthogonality principle, we find
\begin{eqnarray} \label{mmse-filter}
\Hm_{\rm mmse} & = & \frac{\rho}{M} \Hm^\herm \left [\Id + \frac{\rho}{M} \Hm \Hm^\herm\right]^{-1}
= \left [ \Hm^\herm \Hm + \frac{M}{\rho} \Id \right ]^{-1}
\Hm^\herm.
\end{eqnarray}
A standard calculation \cite{Ver} yields the SINRs $\gamma_k$ of
the resulting set of virtual parallel channels in the form
\begin{eqnarray} \label{sinr-mmse}
\gamma_k & = & \frac{\rho}{M} \hv_k^\herm \left [ \Id + \frac{\rho}{M} \Hm_k \Hm_k^\herm \right ]^{-1} \hv_k
= \frac{1}{\left [\left (\Id + \frac{\rho}{M} \Hm^\herm \Hm
\right )^{-1} \right ]_{kk}} - 1,
\end{eqnarray}
where $\Hm_k$ denotes the $N \times (M-1)$ matrix obtained by
removing the $k^{\text{th}}$ column, $\hv_k$, from $\Hm$.

\section{Diversity-Multiplexing Tradeoff} \label{DMT-analysis}

A compact and convenient characterization of the tradeoff between
rate and block-error probability of MIMO quasi-static fading channels in the
high-SNR regime is provided by the DMT
introduced by \cite{ZheTse}. Consider
a family of space-time coding systems, each of which operates at SNR
$\rho$ with rate $R(\rho)$ and error probability $P_e(\rho)$. We say that this family achieves
multiplexing gain $r$ and the diversity gain $d$ (i.e., the point $(r,d)$ on the DMT plane) if
\[ \lim\limits_{\rho \rightarrow \infty} \frac{R(\rho)}{\log \rho}= r, \;\;
 \lim\limits_{\rho \rightarrow \infty} \frac{\log P_e(\rho)}{\log \rho} = -d.\]
 The latter relation is written briefly as $P_e(\rho) \; \doteq \; \text{SNR}^{-d}$ in the exponential
equality notation of \cite{ZheTse}.

The optimal DMT is the best possible error probability exponent
$d^*(r)$ achievable by any space-time scheme at multiplexing gain $r$.
The standard theory of $\epsilon$-capacity \cite{VerHan} readily yields that $d^*(r)$
is equal to the negative $\rho$-exponent of the information outage probability (\ref{eq:outage}).
For the space-time channel in \eqref{eq:mimo}, $d^*(r)$ is given by the piecewise linear function interpolating
the points $(r, d)$ with coordinates
\[ r = k, \;\;\; d = (M-k)(N-k) \]
for $k = 0,1,\ldots,\min\{M,N\}$, and is zero for $r > \min\{M,N\}$ \cite{ZheTse}.

While $d^*(r)$ is achievable under the optimal receiver (a) in Fig.\ref{block-diag},
the following result characterizes the DMT of the MIMO channel in
\eqref{eq:mimo} under schemes (b) and (c), when the linear receiver is either the ZF
or the MMSE receiver defined above:

\begin{thm} \label{th:dmt-lin}
The DMT of the $M$-transmit, $N$-receive i.i.d. Rayleigh MIMO
channel with $N \geq M$, constrained to use Gaussian codes under either MMSE or ZF linear receivers is given by \footnote{Note: $(x)^+ \eqdef \max\{x,0\}$.}
\begin{equation} \label{dmt-lin}
d^*_{\rm lin}(r) = (N-M+1) \left (1 - \frac{r}{M} \right )^+,
\end{equation}
for both the cases of coding across antennas or pure spatial multiplexing.
\end{thm}

{\bf Proof.} The theorem is proved by developing upper and lower
bounds on $P_{\rm out}^{\text{lin}}(R,\rho)$ for the MMSE receiver in the
configuration (b) of the block diagram of Fig.~\ref{block-diag}. A simple upper bound on the
outage probability for the ZF receiver extends immediately the result
to this case. For configuration (c) the result follows as an immediate corollary.


{\bf  Lower bound on the outage exponent.}
Let $\lambda_{\min}(\Am)$ and $\lambda_{\max}(\Am)$ denote the minimum
and maximum eigenvalues of a Hermitian symmetric matrix $\Am$, and
$\lambda_1 \leq \lambda_2 \leq \cdots \leq \lambda_M$ denote the
ordered eigenvalues of the $M \times M$ Wishart matrix
$\Hm^\herm \Hm$, with joint pdf given by \cite{Tel}
\begin{equation} \label{eq:eig_val_wishart_distrib}
p(\lambdav) = K_{M,N} \prod_{i=1}^{M} \lambda_i^{N-M} \cdot
\prod_{i<j} \left( \lambda_i - \lambda_j \right)^2 \exp \left ( -
\sum_{i=1}^{M} \lambda_i \right ),
\end{equation}
where $K_{M,N}$ is a normalization constant and we have assumed $M \leq N$.

Using (\ref{sinr-mmse}), we can write the mutual information
with Gaussian coding across the antennas and the MMSE receiver as
\begin{eqnarray} \label{mmse-info}
I_{\rm mmse}(\Hm) & = & - \sum_{k=1}^{M} \log \left( \left [\left
(\Id + \frac{\rho}{M} \Hm^\herm \Hm \right )^{-1} \right ]_{kk}
\right).
\end{eqnarray}
Since the function $- \log (\cdot)$ is convex, using Jensen's inequality we have
\begin{eqnarray*}
I_{\rm mmse}(\Hm) & \geq & - M \log \left ( \frac{1}{M}
\sum_{k=1}^{M}  \left [ \left (\Id +
\frac{\rho}{M} \Hm^\herm \Hm \right )^{-1} \right ]_{kk} \right )\\
&=& - M \log \left ( \frac{1}{M} \text{Tr} \left[ \left (\Id +
\frac{\rho}{M} \Hm^\herm \Hm \right )^{-1}
\right] \right )\\
&=& - M \log \left ( \frac{1}{M} \sum_{k=1}^{M} \frac{1}{1 +
\frac{\rho}{M} \lambda_k} \right ).
\end{eqnarray*}
Using this bound in \eqref{eq:pout_lin} we obtain
\begin{eqnarray} \label{eq:mmse_pout_ubd}
P_{\rm out}^{\text{mmse}}(R,\rho) & \leq & P \left ( \log \left ( \frac{1}{M}
\sum_{k=1}^{M} \frac{1}{1 +
\frac{\rho}{M} \lambda_k} \right ) \geq - \frac{R}{M} \right ) \nonumber \\
& = & P \left ( \frac{1}{M} \sum_{k=1}^{M} \frac{1}{1 +
\frac{\rho}{M} \lambda_k} \geq \rho^{-\frac{r}{M}} \right ),
\end{eqnarray}
where in the last line we let  $R = r \log \rho$.
Finally, we can use the trivial asymptotic upper bound
\begin{equation} \label{eq:MMSE_pout_lbd}
P \left ( \frac{1}{M} \sum_{k=1}^{M} \frac{1}{1 +
\frac{\rho}{M} \lambda_k} \geq \rho^{-\frac{r}{M}} \right ) \;
\dot\leq \; P \left ( \frac{1}{\rho \lambda_1} \geq
\rho^{-\frac{r}{M}} \right )
\end{equation}
First, we notice that the asymptotic outage probability upper bound in
the RHS of (\ref{eq:MMSE_pout_lbd}) vanishes only if $r/M < 1$.
Hence, the outage exponent lower bound is zero for $r/M \geq 1$.
When $r/M < 1$, we can write
\begin{eqnarray} \label{aris1}
P \left ( \lambda_1 \leq \rho^{\frac{r}{M} - 1} \right ) & = &
\int_0^{\rho^{\frac{r}{M} - 1}} d\lambda_1 \prod_{i=2}^M
\left [ \int_{\lambda_1}^\infty d\lambda_i \right ] p(\lambdav) \nonumber \\
& \leq & \int_0^{\rho^{\frac{r}{M} - 1}} d\lambda_1 \prod_{i=2}^M
\left [ \int_0^\infty d\lambda_i \right ] p(\lambdav) \nonumber \\
& = & \int_0^{\rho^{\frac{r}{M} - 1}} p_1(\lambda_1) d\lambda_1 \nonumber \\
& = & \kappa_1 \rho^{(N - M + 1)(r/M - 1)},
\end{eqnarray}
where $\kappa_1$ is a constant and where we have used the
well-known fact \cite{Tel} that the marginal pdf of $\lambda_1 = \lambda_{\min}(\Hm^\herm \Hm)$,
denoted by $p_1(\lambda)$ in (\ref{aris1}),
satisfies $p_1(\lambda) \propto \lambda^{N-M}$ for small argument $\lambda \ll 1$.
The resulting outage exponent lower bound is
\begin{equation} \label{MMSE-exponent-LB1}
d^*_{\rm mmse}(r) \geq (N - M + 1) \left( 1 - \frac{r}{M} \right)^+.
\end{equation}
The same result can be obtained by following the by-now
standard technique of \cite{ZheTse} based on the change of variable
$\lambda_i = \rho^{-\alpha_i}$, integrating the resulting pdf of
$\alpha_1,\ldots,\alpha_M$ over the outage region and applying
Varadhan's lemma \cite{ZheTse}.


{\bf Upper bound on the outage exponent.} Using the concavity and
the monotonicity of the $\log (\cdot)$ function, we obtain from
\eqref{mmse-info} and Jensen's inequality that
\begin{eqnarray} \label{ziocane}
I_{\rm mmse}(\Hm) \leq M \log \left( \frac{1}{M} \sum_{k=1}^{M}
\frac{1}{\left [ \left (\Id + \rho \Hm^\herm \Hm \right )^{-1}
\right ]_{kk}} \right).
\end{eqnarray}
Consider the decomposition $\Hm^\herm \Hm = \Um^\herm \Lambdam
\Um$, where $\Um$ is unitary and $\Lambdam$ is a diagonal matrix
with the eigenvalues of $\Hm^\herm \Hm$ on the diagonal. Defining
$\uv_k$ to be the $k^{\text{th}}$ column of $\Um$ and $\ev_k$ to
be the column vector that has a one in the $k^{\text{th}}$
component and zeros elsewhere, we have that
\begin{eqnarray*}
\left [(\Id + \rho \Hm^\herm \Hm)^{-1} \right ]_{kk} &=& \ev_k^\herm \Um^\herm \left ( \Id + \rho \Lambdam \right )^{-1} \Um \ev_k\\
&=& \uv_k^\herm \left ( \Id + \rho \Lambdam \right )^{-1} \uv_k\\
&=& \sum_{\ell = 1}^{M} \frac{|u_{\ell k}|^2}{1 + \rho \lambda_\ell}.
\end{eqnarray*}
Hence, the term inside the logarithm in (\ref{ziocane}) can be upperbounded as
\begin{eqnarray}
\frac{1}{M} \sum_{k=1}^{M} \frac{1}{\left [ \left (\Id + \rho
\Hm^\herm \Hm \right )^{-1} \right ]_{kk}} & = &
\frac{1}{M} \sum_{k=1}^{M} \frac{1}{\sum_{\ell = 1}^{M} \frac{|u_{\ell k}|^2}{1 + \rho \lambda_\ell} }  \nonumber \\
& = & \frac{1}{M} \sum_{k=1}^{M} \frac{1}{ \frac{|u_{1 k}|^2}{1+\rho \lambda_1} \left[ 1 + \sum_{\ell = 2}^{M} \frac{|u_{\ell k}|^2}{|u_{1 k}|^2} \frac{1 + \rho \lambda_1}{1 + \rho \lambda_\ell} \right] } \nonumber \\
& \leq & (1 + \rho \lambda_1) \frac{1}{M} \sum_{k=1}^{M}
\frac{1}{|u_{1 k}|^2}
\end{eqnarray}
Let $\mathcal{A}$ denote the event $\left\{ \frac{1}{M}
\sum_{k=1}^{M} \frac{1}{|u_{1 k}|^2} \leq c \right\}$, where $c$
is some constant (independent of $\rho$). We have that
\begin{eqnarray} \label{dioffa}
P_{\rm out}^{\text{mmse}} (R,\rho) & \geq & P \left ( \mathcal{A} \right )
P \left ( \left. \log \left( (1 + \rho \lambda_1) \frac{1}{M}
\sum_{k=1}^{M} \frac{1}{|u_{1 k}|^2} \right) \leq \frac{R}{M}
\right | \mathcal{A}\right ) \nonumber \\
&\geq& P \left ( \mathcal{A} \right ) P \left ( \log \left( (1 +
\rho \lambda_1) c \right) \leq \frac{R}{M} \right ) \nonumber \\
& \doteq & P \left ( \log \left( 1 + \rho \lambda_1 \right ) \leq
\frac{r}{M} \log \rho  \right )
\end{eqnarray}
where the last exponential equality holds if
$P \left ( \mathcal{A} \right )$ is a $O(1)$ non-zero term, i.e., it is a
constant with respect to $\rho$ bounded away from zero.
This is indeed the case, as shown rigorously in Appendix \ref{appendix-A}.

It is immediate to check that the last line of (\ref{dioffa}) is asymptotically equivalent to
\eqref{eq:MMSE_pout_lbd}. Therefore, applying the same argument as in (\ref{aris1}) we find
that the upper bound on the outage probability exponent coincides with the previously found lower bound.

The proof of Theorem \ref{th:dmt-lin} is completed by
observing that in the case of the ZF receiver a lower bound on the SINR $\gamma_k$
is readily obtained from the inequality
\[ \left [(\Hm^\herm \Hm)^{-1} \right ]_{kk} \leq \lambda_{\max}[(\Hm^\herm \Hm)^{-1}] = \frac{1}{\lambda_{\min}(\Hm^\herm \Hm)} = \frac{1}{\lambda_1}, \]
that holds for all $k = 1,\hdots,M$.
Using this in the mutual information expression for the ZF receiver with coding across the antennas we obtain
\begin{eqnarray} \label{eq:zf_pout_ubd}
P_{\rm out}^{\rm zf}(R,\rho) & \leq & P \left ( \log (1 + \rho
\lambda_1) \leq \frac{r}{M} \log \rho \right )
\end{eqnarray}
Noticing that (\ref{eq:zf_pout_ubd}) coincides with the asymptotic lower bound (\ref{dioffa}) for the MMSE receiver, and that
the MMSE receiver maximizes the mutual information over all linear receivers,
under Gaussian inputs and the system assumptions made here,  we immediately obtain that the ZF also achieves the outage exponent
$d^*_{\rm lin}(r)$ given in (\ref{dmt-lin}).

Finally, as far as spatial multiplexing is concerned (no coding across the antennas),
it is clear from (\ref{eq:pout_lin}) and (\ref{eq:pout_sp_mult}) that, for any linear receiver $\Gm$, $P_{\rm out}^{\text{lin}}(R,\rho) \leq
P_{\rm out}^{\text{sp mult}}(R,\rho)$. On the other hand, it is immediate to show that spatial multiplexing achieves the same
DMT (\ref{dmt-lin}).  Details are trivial, and then are omitted. \hfill $\square$

\subsection{Discussion and numerical results} \label{discussion}

Theorem \ref{th:dmt-lin} shows that, in terms of DMT, there is no
advantage in using interleaving and coding across the antennas
when a linear receiver is used in order to spatially separate the
transmitted symbols. In order words, the linear receiver front-end
kills the {\em transmit} diversity gain offered by the MIMO channel.
In fact, the DMT $(N-M+1) \left( 1 - \frac{r}{M}
\right)^+$ of Theorem \ref{th:dmt-lin} has the
following intuitive interpretation: this coincides with the DMT
of a SIMO (Single-Input, Multiple-Output) channel (receiver diversity only)
with $N-M+1$ receive antennas, used at a rate $R/M$.

This fact shows also that the channel gains of the virtual parallel
channels are {\em strongly} statistically dependent. For example, it
is well-known that the ZF receiver applied to a $M \times N$ channel
with $N \geq M$ and i.i.d. Rayleigh fading yields channel gains
$\gamma_k = \frac{1}{\left [(\Hm^\herm \Hm)^{-1} \right ]_{kk}}$
that are marginally distributed as central Chi-squared random
variables with $2(N-M+1)$ degrees of freedom \cite{WinSalGit}. If
the gains $\gamma_1,\ldots,\gamma_M$ were statistically independent,
by coding across the antennas we would obtain the DMT of the
parallel independent channels, given by \cite{TseVis}
\[ d^*_{\rm parallel,i.i.d.} (r) \geq (N-M+1) \left( M - r \right)^+,  \]
which is much larger than the DMT given by Theorem \ref{th:dmt-lin}.
In contrast, the channel gains in the regime of high SNR
are essentially dominated by the minimum eigenvalue of the matrix $\Hm^\herm  \Hm$
and therefore are strongly correlated: if one subchannel is in deep fade, they are all in deep fade with
high probability. This is the reason why coding across the transmit antennas does not buy any improvement in terms of
DMT with respect to simple spatial multiplexing.\footnote{This also show that the assumption that the $\gamma_k$'s are i.i.d., made in \cite{HedNos},
is incompatible with the  final result of that paper on the diversity of the ZF receiver.}

\begin{figure}[htbp]
\begin{center}
\includegraphics[width=15cm]{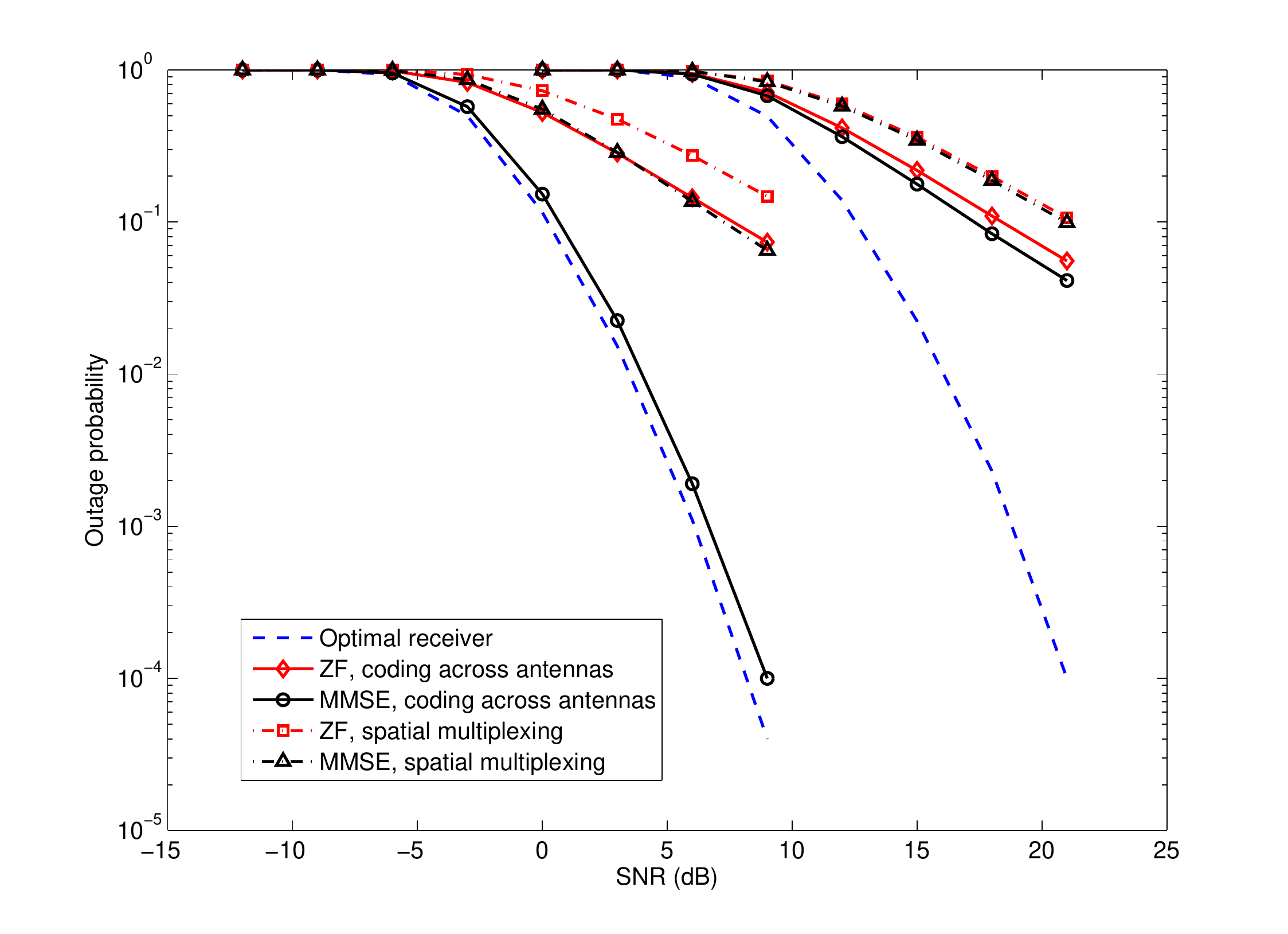}
\caption{Outage probabilities of ZF and MMSE receivers, $2 \times 2$
i.i.d. Rayleigh channel, $R = 1$ and $5$ bpcu.}
\label{fig:pout_mmse_zf_1_5}
\end{center}
\end{figure}

Having said so, we should also remark that the picture about linear receivers is not totally
grim as it may appear from the high-SNR DMT analysis. Indeed,
coding across antennas yields a very significant performance
advantage with the linear MMSE receiver at fixed and not too large
rate (notice that fix rate $R$ corresponds to
the case of zero multiplexing gain, $r = 0$.)
In order to illustrate these claims,  we provide simulations results for the following outage
probabilities under i.i.d. Rayleigh fading:
\begin{itemize}
\item MIMO outage probability \eqref{eq:outage} with input
covariance $(\rho/M) \Id$ (scheme (a) in Fig.\ref{block-diag});
\item outage probability \eqref{eq:pout_lin} with ZF and MMSE receivers with
coding across antennas ((scheme (b) in Fig.\ref{block-diag}));
\item outage probability \eqref{eq:pout_sp_mult} with ZF and MMSE
receivers under pure spatial multiplexing, i.e., without coding
across antennas (scheme (c) in Fig.\ref{block-diag}).
\end{itemize}
Fig.~\ref{fig:pout_mmse_zf_1_5} shows the corresponding plots at rates $R = 1$ and $5$ bits per channel use (bpcu).

Several interesting observations can be drawn from this figure.
We observe that while at high rates the MMSE with coding across
antennas behaves as predicted by the DMT analysis, the behavior at
low rates is in stark contrast to the asymptotic result (this fact was also noticed in \cite{HedNos}).
In fact, the MMSE exhibits an apparent ``full diversity'' behavior at small rate (e.g., $R = 1$
bpcu in Fig.~\ref{fig:pout_mmse_zf_1_5}). In contrast, the behavior of the ZF
receiver is accurately predicted by the asymptotic analysis at all rates. This remarkable behavior of the MMSE receiver
is explained through an approximate analysis in Section~\ref{sec:MMSE_finite_rate}.

From Fig.~\ref{fig:pout_mmse_zf_1_5} we observe also that coding
across antennas does achieve an advantage over spatial
multiplexing. For the MMSE receiver operating at small rates the
advantage is very significant, and corresponds to the diversity
advantage discussed above. At high rates the advantage is moderate and
consists only of a horizontal shift (dB gain) of the error curve, not in a steeper slope.

\section{MMSE receiver with coding across antennas} \label{sec:MMSE_finite_rate}

\begin{figure}[htbp]
\begin{center}
\includegraphics[width=12cm]{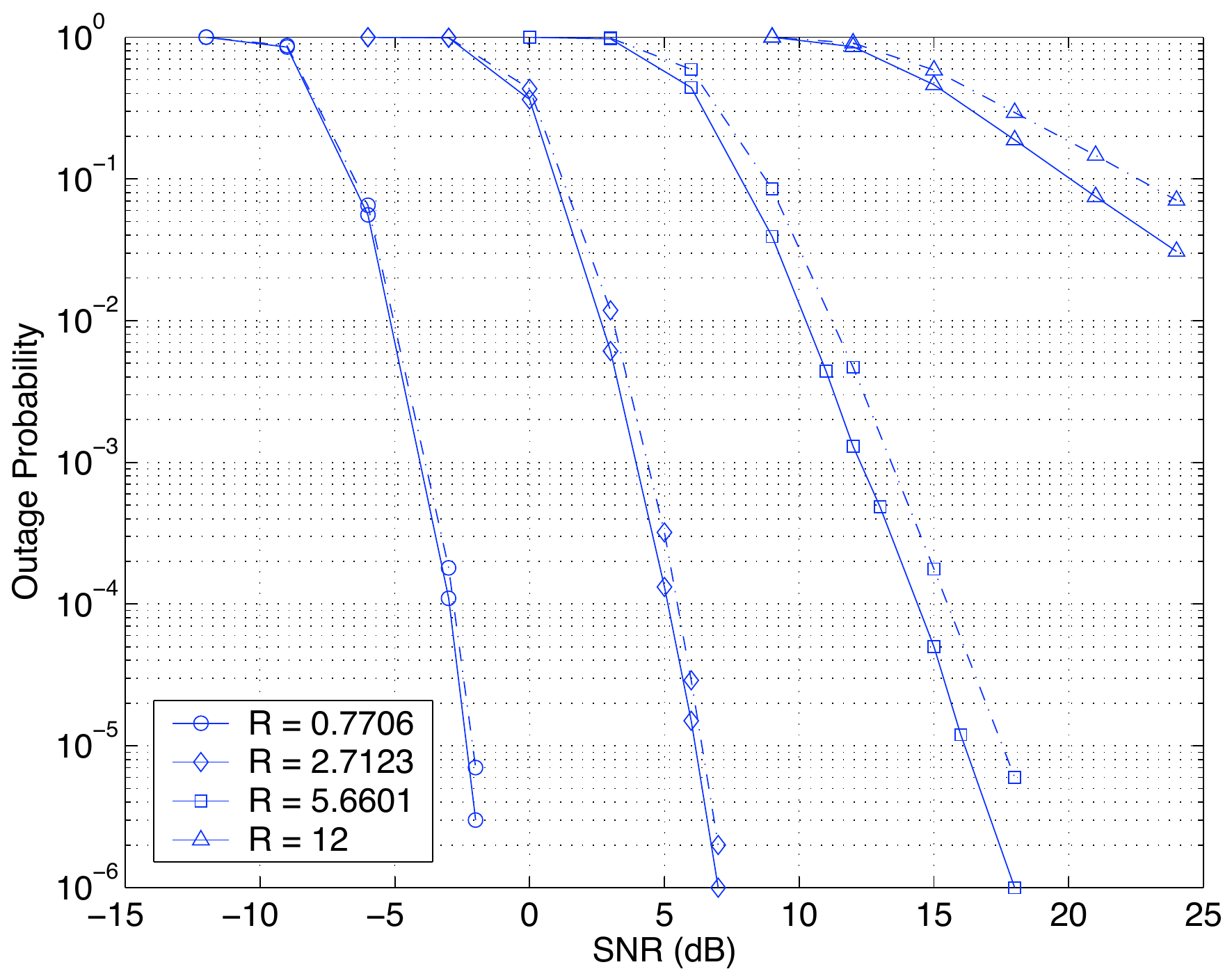}
\caption{Diversity of the MMSE receiver with joint spatial encoding:
solid lines represent the outage probability in \eqref{eq:pout_lin}
and the dash-dot lines represent the corresponding upper bounds
\eqref{eq:mmse_pout_ubd}. $M = N = 4$, rates $R$ are in bpcu.}
\label{fig:MMSE_finite_rate}
\end{center}
\end{figure}

The difference between the performances of the ZF and MMSE
receivers is best explained by comparing their corresponding upper
bounds on outage probability in \eqref{eq:zf_pout_ubd} and
\eqref{eq:mmse_pout_ubd}. While only the minimum eigenvalue
appears in the ZF case in \eqref{eq:zf_pout_ubd}, all eigenvalues
play a role in the case of the MMSE receiver in
\eqref{eq:mmse_pout_ubd}. Although at asymptotically high SNR and
high coding rates the minimum eigenvalue dominates (and therefore
determines the corresponding DMT), the other eigenvalues appear to
be relevant at lower rates and provide higher effective diversity
for the MMSE receiver. In order to substantiate this intuition, we
compare in Fig.~\ref{fig:MMSE_finite_rate} the outage probability
of the MMSE receiver with coding across antennas for the case $M =
N = 4$ with the corresponding upper bound in
\eqref{eq:mmse_pout_ubd}.
The upper bound is found to be very accurate
across a wide range of rates and SNRs. The particular choice of
rates for this plot will be made clear in the sequel, where we
analyze the high SNR behavior of the outage probability upper
bound \eqref{eq:mmse_pout_ubd}.

Define ${\mathfrak T}_k \triangleq \frac{1}{1 + \frac{\rho}{M}
\lambda_k}$ and $\mathfrak{T} \triangleq M 2^{-\frac{R}{M}}$. We
use a change of variables $\lambda_k = \rho^{-\alpha_k}$, where
$\alpha_k$ denotes the {\em level of singularity} of the
corresponding eigenvalue \cite{ZheTse}. For ease of analysis we
make the assumption that the channel eigenvalues fall into one of
the following two categories: \bit
\item $\alpha_k < 1$, i.e., $\lambda_k$ is ``much larger'' than the inverse SNR $1/\rho$:
in this case, $\mathfrak{T}_k \rightarrow 0$ as $\rho \rightarrow
\infty$.
\item $\alpha_k > 1$, i.e., $\lambda_k$ is ``much smaller'' than $1/\rho$:
in this case, $\mathfrak{T}_k \rightarrow 1$ as $\rho \rightarrow
\infty$. \eit Recall that the $\{\alpha_i\}$ are ordered according
to $\alpha_1 \geq \cdots \geq \alpha_M$. Suppose that the rate $R$
is such that $m - 1 < \mathfrak{T} \leq m$, for some integer $m =
1,2,\hdots,M$, i.e.,
\begin{equation} \label{mR}
M \log \frac{M}{m} \leq R < M \log \frac{M}{m-1}.
\end{equation}
For all $i = 1,\ldots,M$ define the event
\begin{equation}
\Ec_i = \{ \alpha_1,\ldots,\alpha_i > 1\} \cap
\{\alpha_{i+1},\ldots,\alpha_M < 1 \}.
\end{equation}
Then, for large $\rho$, the following approximation holds
\begin{eqnarray} \label{key-approx}
\left\{ \sum_{k=1}^{M} \mathfrak{T}_k \geq \mathfrak{T} \right\} &
\approx &
\bigcup_{i=m}^M  \left \{ \alpha_1,\ldots,\alpha_i > 1 \right \} \cap \left \{\alpha_{i+1},\hdots,\alpha_M < 1 \right\} \nonumber \\
& = & \mathcal{E}_m \cup \mathcal{E}_{m + 1} \cup \cdots \cup
\mathcal{E}_M.
\end{eqnarray}
In the above approximation we are neglecting the cases where the
eigenvalues take on values that are comparable with $1/\rho$, and
therefore contribute to the sum $\sum_{k=1}^{M} \mathfrak{T}_k$ in \eqref{eq:mmse_pout_ubd}
by a quantity between 0 and 1. It can be expected that as $\rho
\rightarrow \infty$, the probability of such intermediate values
decreases, and our approximation becomes tight.

Using the union bound, we find an approximate upper bound on \eqref{eq:mmse_pout_ubd} given by
\begin{equation} \label{eq:union_bd}
P \left( \sum_{k=1}^{M} \mathfrak{T}_k \geq \mathfrak{T} \right) \
\lesssim \ \sum_{i=m}^{M} P (\mathcal{E}_i).
\end{equation}
Defining $P (\mathcal{E}_i) \doteq \rho^{-\tilde{d}_i(R)}, \
i=1,\hdots,M$, using the joint pdf of the $\alpha_k$'s, given by
\cite{ZheTse}
\begin{eqnarray*}
p(\alphav) & = & K_{M,N} \left[ \log (\rho) \right]^{M}
\prod_{i=1}^{M} \rho^{-(N-M+1)\alpha_i} \prod_{i<j} \left(
\rho^{-\alpha_i} - \rho^{-\alpha_j} \right)^2
\exp \left ( - \sum_{i=1}^{M} \rho^{-\alpha_i} \right ) \\
& \doteq & \left[ \prod_{i = 1}^{M} \rho^{-(2i-1+N-M)\alpha_i}
\right] \exp \left ( - \sum_{i=1}^{M} \rho^{-\alpha_i} \right ),
\end{eqnarray*}
and applying Varadhan's lemma as in \cite{ZheTse}, we obtain
\begin{eqnarray} \label{eq:d_tilde}
\tilde{d}_{i} (R) & = & \inf\limits_{\begin{array}{c}
\alpha_j > 1 \ \forall \ j \leq i \nonumber \\
\alpha_j < 1 \ \forall \ j > i\\
\alpha_j \geq 0 \ \forall \ j \end{array}} \sum_{j=1}^{M}
(2j-1+M-N)
\alpha_j \nonumber \\
&=& \sum_{j=1}^{i} (2j-1+M-N) \times 1  + \sum_{j=i+1}^{M} (2j-1+M-N) \times 0 \nonumber \\
&=& i(i+N-M).
\end{eqnarray}
From \eqref{eq:union_bd} and \eqref{eq:d_tilde}, we eventually
conclude that
\[ P \left( \sum_{k=1}^{M} \mathfrak{T}_k \geq \mathfrak{T} \right) \ \dot\lesssim \ P(\mathcal{E}_m). \]
This yields the diversity of the MMSE receiver with spatial
encoding at a finite rate $R$ as
\begin{equation} \label{eq:mmse_finite_rate_diversity}
d_{\rm mmse} (R) \ \approx \ m(m + N - M).
\end{equation}
In particular, when $M = N$, $d_{\rm mmse} (R) \ \approx \ m^2$
where $m$ and $R$ are related by (\ref{mR}).

To illustrate the effectiveness of the above approximation,
consider the plots in Fig.~\ref{fig:MMSE_finite_rate} for the case
$M = N = 4$. The coding rates are $R = 0.7706, \ 2.7123, \ 5.6601$
and $12$ bpcu, corresponding to $\mathfrak{T} = 3.5, \ 2.5, \ 1.5$
and $0.5$ respectively. The diversities $16, \ 9, \ 4$ and $1$
predicted by the analysis in \eqref{eq:mmse_finite_rate_diversity}
well approximate the measured slopes (for high SNR) of the outage
curves, that are $15.15, \ 10.69, \ 5.55$ and $1.3$ in the $\log
P_{\rm out}^{\text{mmse}}(R,\rho)$ vs. $\log \rho$ chart observed in
Fig.~\ref{fig:MMSE_finite_rate}.

\section{Outage probability of linear receivers in the large antenna regime} \label{sec:MMSE_MI_distrib}

\begin{figure}[htbp]
\begin{center}
\includegraphics[width=15cm]{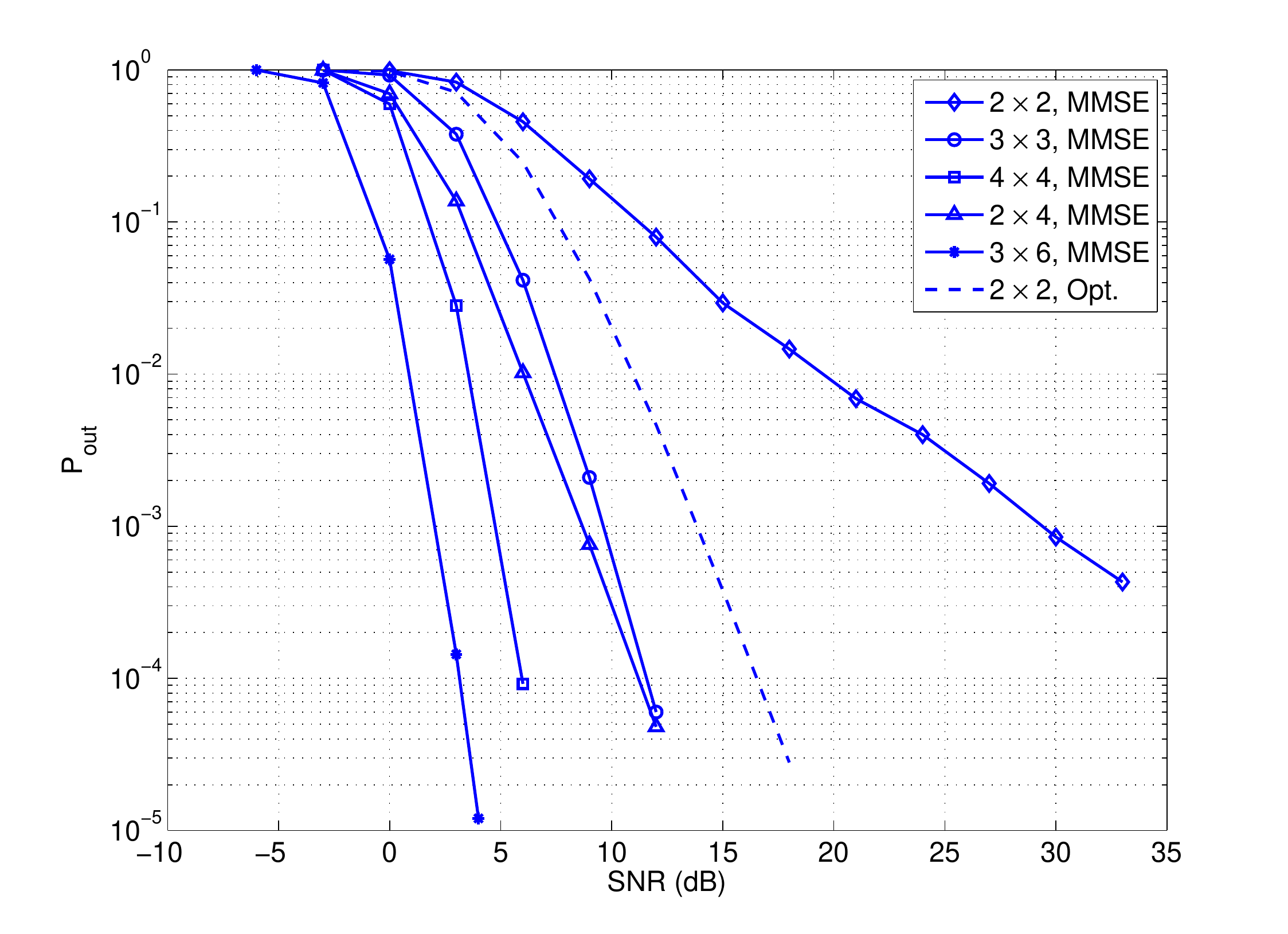}
\caption{Comparing the outage probability of optimal and MMSE
receivers, $R = 3$ bpcu.} \label{fig:increase_antennas}
\end{center}
\end{figure}

In order to motivate this section, consider the following system
design issue: for a given target spectral efficiency, block-error
rate, operating SNR, and receiver computational complexity
(including power consumption, VLSI chip area etc.) how many antennas
do we need at the transmitter and receiver? Consider the outage
probability curves of Fig.~\ref{fig:increase_antennas} and suppose
that we wish to achieve a rate of $R = 3$ bpcu with block-error rate
of $10^{-3}$ at SNR not larger than 15 dB. With $M = N = 2$ antennas
this target performance is achieved by an optimal receiver, but is
not achieved by the MMSE receiver. However, with $M = 2, N= 4$ or $M
= N = 3$ the target performance is achieved also by the MMSE
receiver. It turns out that, in some cases, adding antennas may be
more convenient than insisting on high-complexity receiver
processing.

It is therefore interesting to analyze the outage probability of
a linear  receiver with coding across the
antennas in the regime of fixed SNR $\rho$ and rate $R$. This
analysis is difficult due to the fact that, for finite $M, N$, the
joint distribution of the channel SINRs $\{\gamma_k\}$ in
(\ref{parallel-ch}) escapes a closed-form expression. This problem
can be overcome by considering the system in the limit of a large number of antennas.
Specifically, we will show that the mutual information for
the linear MMSE and ZF receivers becomes asymptotically Gaussian. Therefore, the outage
probability for large but finite dimensions and fixed SNR can be accurately
approximated by a Gaussian cdf  with appropriate mean and
variance, that we shall give in closed form.

In the next subsection we will discuss the methodology used to show
the asymptotic Gaussianity of the mutual information. The method is
general and applies to both MMSE and ZF linear receivers.
Subsequently, in Section~\ref{sec:Joint_cum_moments12} we will
calculate the first and second cumulant moments of the SINR for the
MMSE and ZF receivers, which suffice to characterize the mutual information
limiting distribution.

\subsection{Asymptotic Gaussianity of the mutual information}
\label{sec:gaussianity_ofMI}

The mutual information at the output of a linear
receiver with $M$ transmit and $N \geq M$ receive
antennas and coding across the antennas is given by
\begin{equation} \label{IN}
I_N \triangleq \sum_{k=1}^M \log \left ( 1 + \gamma_k \right )
\end{equation}
with $\gamma_k$ given by (\ref{sinr-mmse}) for the MMSE case
and by (\ref{sinr-zf}) for the ZF case. In the following, we
fix the ratio $\beta = M/N \leq 1$ and
consider the limit for large $N$ and the ``fluctuations'' around this limit.
In order to prove the asymptotic Gaussianity
of these fluctuations, we will need to analyze the
characteristic function of the mutual information, given by
\begin{equation}\label{eq:mg_function_def}
\Phi_N(\omega) \triangleq \EE \left[ e^{j \omega I_N} \right].
\end{equation}
We start by considering the {\em cumulant generating function}
\cite{Men}, defined as
\begin{equation}\label{eq:cum_mg_function_def}
\phi_N(\omega) \triangleq \log(\Phi_N(\omega)) = \sum_{n=1}^\infty
\frac{(j \omega)^n}{n!} {\cal C}_n,
\end{equation}
where the coefficient ${\cal C}_n$ is the $n$-th cumulant moment of
the mutual information. In general, the {\em joint cumulant} of $m$
random variables $X_1,\hdots,X_m$ is defined as
\[ \EE_c (X_1;\hdots ;X_m) \triangleq \sum_\pi (|\pi|-1)! (-1)^{|\pi|-1} \prod_{B \in \pi}
\EE\left [  \prod_{i \in B} X_i \right ], \] where $\pi$ runs
through all partitions of $\{1,\hdots,m \}$, $|\pi|$ denotes the
number of blocks in $\pi$ and $B$ runs through the list of all
blocks of $\pi$.
We will call the above moment irreducible,
with respect to the random variables $X_1,\ldots,X_m$,
when in each argument of the cumulant moment only one random
variable $X_i$ appears. By contrast a {\em reducible} cumulant
moment with respect to the same random variables has arguments
containing mixed products of these random variables.
In general, an $n$-order reducible cumulant moment can be written
in terms of a sum of products of irreducible cumulant moments,
with each term in the sum having moments with
order summing up to $n$.

The $n^{\text{th}}$ cumulant moment $\Cc_n$ of a random variable
$X$ is defined to be
\[\Cc_n \triangleq \EE_c(\underbrace{X; \hdots ; X}_{n \text{ times}}). \]
For example, the first few cumulant moments of $X$ are
\begin{eqnarray}\label{eq:cum_def}
{\cal C}_1 &=& \EE[X]\;\;\;\;\;\; \mbox{mean} \\ \nonumber %
{\cal C}_2 &=& \text{Var}[X] \;\;\;\;\;\; \mbox{variance}\\ \nonumber %
{\cal C}_3 &=& \text{Sk}[X] \;\;\;\;\;\; \mbox{skewness}
\end{eqnarray}
The probability density of $I_N$ can be expressed in
terms of (\ref{eq:mg_function_def}) and
(\ref{eq:cum_mg_function_def}) as follows
\begin{eqnarray}\label{eq:IFFT_I_N}
p (y) &=& \frac{1}{2\pi} \int_{-\infty}^\infty e^{-j \omega y}
\Phi_N(\omega) d\omega \nonumber \\
&=& \frac{1}{2\pi} \int_{-\infty}^\infty \exp \left ( -j \omega
\left( y - {\cal C}_1 \right) - \frac{\omega^2}{2} {\cal C}_2
+\sum_{n>2} \frac{(j \omega)^n}{n!} {\cal C}_n \right )  d\omega.
\end{eqnarray}
In Section~\ref{sec:Joint_cum_moments12}, we will show that in the
limit of large $N$ and $M = \beta N$ with $\beta \leq 1$,
\begin{eqnarray}\label{eq:asympt_C_1}
{\cal C}_1 & = & m_1 + o(1)  \\
\label{eq:asympt_C_2}
{\cal C}_2 & = & \sigma^2 + o(1)
\end{eqnarray}
where $m_1 = M c_{10} + c_{11}$, and where $c_{10}$, $c_{11}$ and
$\sigma^2$ are constants independent of $N$ for which we give
closed-form expressions for both MMSE and ZF cases. In
Appendix \ref{app:asympt_cum} we will also show that
all higher-order cumulants of the mutual information
asymptotically vanish for large $N$. Therefore,
$\phi_N(\omega)$ is a quadratic function of $\omega$ with corrections that vanish as
$N \rightarrow \infty$. As a result the mutual information is asymptotically Gaussian, i.e.
\begin{equation}\label{eq:MI_Gaussian}
 \frac{I_N - m_1}{\sigma} \; \stackrel{d}{\rightarrow} \; {\cal N}(0,1).
\end{equation}
This follows directly from (\ref{eq:IFFT_I_N}) by setting
$y = z + m_1$ and taking the large $N$ limit
\begin{eqnarray}\label{eq:IFFT_CLT}
p(z) &=& \lim_{N\rightarrow\infty} \frac{1}{2\pi}
\int_{-\infty}^\infty \exp \left ( -j \omega  z - \frac{\omega^2}{2}
\sigma^2
+o(1)\right )  d\omega \\ \nonumber %
&=& \frac{1}{{\sqrt{2\pi \sigma^2}}}  e^{-\frac{z^2}{2\sigma^2}}.
\end{eqnarray}
Before moving on to the proofs, we would like to comment on
the nature of this result. This states that the probability $P( |I_N - m_1| > z)$
approaches a Gaussian probability for sufficiently large $N$ and
{\em  fixed} distance $z$ of the mutual information from its mean.
This is quite different from stating that for fixed $N$ the mutual information
distribution falls off like a Gaussian random variable for any $z$ and $\rho$.
As a matter of fact, for {\em fixed} $N$ and large enough SNR this Gaussian approximation is
no longer valid, since the higher-order cumulants will no longer be small.

It is also worth pointing out here that the variance of $I_N$ is
$O(1)$  (a {\em finite} constant) for large $N$.
This another manifestation of the fact that the SINRs of the parallel channels $\{\gamma_k\}$ are
strongly correlated, in agreement with the outage analysis of
previous sections. In contrast, if they were independent, or nearly
independent, the variance would be roughly linear in $N$, as the
central limit theorem would suggest. This fact is in line with
the well-known behavior of the mutual information $\log\det(\Id + \frac{\rho}{M} \Hm\Hm^\herm )$
under the optimal receiver  \cite{Hochwald2002_MultiAntennaChannelHardening,
Moustakas2003_MIMO1,Smith2002_OnTheGaussianApproximationToTheCapacityOfWirelessMIMOSystems,
Hachem2007_NewApproachGaussianMIMO}, where again the variance
is $O(1)$ for large $N$, indicating the  strong correlation among the eigenvalues of $\Hm^\herm \Hm$.

\subsection{Joint cumulant moments of the SINRs of order 1 and 2}
\label{sec:Joint_cum_moments12}

Our goal is to calculate the cumulant moments of $I_N$. Since $I_N$
consists of a sum of mutual informations of the virtual channels
(see (\ref{IN})), the $n^{\text{th}}$ cumulant moment of $I_N$ can
be written as
\begin{eqnarray}\label{eq:cum_product_def}
{\cal C}_n = \sum_{k_1,\ldots
k_n=1}^M\EE_c[\log(1+\gamma_{k_1});\ldots;\log(1+\gamma_{k_n})].
\end{eqnarray}
The building blocks of the above cumulant moments are the
 joint cumulant moments of the SINRs $\{\gamma_k\}$, i.e.
\begin{eqnarray}\label{eq:cum_product_def3}
\EE_c[\gamma_{k_1};\gamma_{k_2};\ldots;\gamma_{k_n}].
\end{eqnarray}
In fact, by expanding the logarithms in (\ref{eq:cum_product_def})
in Taylor series, we can express (\ref{eq:cum_product_def}) in terms
of (\ref{eq:cum_product_def3}). Even calculating these joint
cumulants amounts generally to a formidable task. However, with the
help of Theorem~\ref{thm:Novikov} (Novikov's theorem) given in
Appendix~\ref{app:novikov_thm} and due to simplifications that occur
in the large $N$ limit, we will show that this computation is
possible. To obtain a feel for the computation, we will first
calculate the first two joint cumulants of $\{\gamma_k\}$ and defer
the proof that the higher-order cumulants vanish sufficiently fast
with $N$ to Appendix \ref{app:asympt_cum}.

\subsubsection{Cumulant moments for the MMSE receiver}\label{sec:cum_moments12_MMSE}

Starting with the case of the MMSE receiver, we
recall from \eqref{sinr-mmse} that the SINR of the $k$-th virtual channel
induced by the MMSE receiver can be written as
\begin{equation}\label{eq:sir_def}
  \gamma_k = \alpha  \hv^\herm_k
  \left[ \Id + \alpha \Hm_k \Hm_k^\herm \right]^{-1} \hv_k
\end{equation}
where $\Hm_k$ is the $N\times (M-1)$ matrix obtained by
eliminating the $k$-th column $\hv_k$ from the channel matrix
$\Hm$, and contains i.i.d. Gaussian elements  $\sim \Cc \Nc
(0,1/N)$ and we have defined for convenience $\alpha = \rho N/M =
\rho/\beta$.

The asymptotic mean of $\gamma_k$ in the limit of large $N$ and $M
= \beta N$ has been calculated in \cite{verdu-shamai99} in the context of
large-system analysis of CDMA with random spreading,
and successively rederived in various ways
(e.g., \cite{Tse1999_LMUReceivers, Tse2000_MMSEFluctuations, Debbah2003_UnitaryAsymptoticallyFreeMatrices}). %
Due to symmetry, the result does not depend on the index $k$. Hence,
without loss of generality we can choose $k = 1$. We have
\begin{eqnarray}\label{eq:mean_sir_gen}
\EE[ \gamma_1] & = & \alpha
  \EE \left [  \frac{1}{N} {\rm Tr} \left ( \left[ \Id  + \alpha
  \Hm_1 \Hm_1^\herm \right]^{-1} \right ) \right].
\end{eqnarray}
The leading order in $N$ of the above trace can be evaluated as
\begin{eqnarray}\label{eq:mean_sir}
  g_1^\mmse(\alpha,\beta) &  = & \lim_{N \rightarrow \infty}  \alpha
  \EE \left [  \frac{1}{N} {\rm Tr} \left ( \left[ \Id  + \alpha
  \Hm_1 \Hm_1^\herm \right]^{-1} \right ) \right] \nonumber \\
  &=& \frac{\alpha}{1 + \frac{\alpha\beta}{1 + g_1^\mmse(\alpha,\beta)}}.
\end{eqnarray}
Solving for $g_1^\mmse(\alpha,\beta)$ in (\ref{eq:mean_sir}), we
obtain
\begin{equation}\label{eq:ave_SIR}
g_1^\mmse(\alpha,\beta) = \frac{1}{2}\left[\alpha (1 - \beta) - 1 +
\sqrt{(\alpha(1 - \beta) - 1)^2 + 4\alpha} \right].
\end{equation}
To be able to calculate the $O(1)$ correction to the mean mutual
information, we need to evaluate the next to leading ($O(1/N)$)
correction to $\EE[\gamma_1]$. The correction follows by noticing that
the term $\beta$ in (\ref{eq:mean_sir}) should be replaced by the aspect ratio of
the matrix $\Hm_1$.  For large but finite $N$, this is equal to
$(M-1)/N = \beta - 1/N$. Therefore, the correction can be
evaluated by replacing $\beta$ by $\beta - 1/N$ in
\eqref{eq:ave_SIR}.
Using the Taylor series expansion, this amounts to computing
\begin{eqnarray}\label{eq:mean_sir__plus_correction}
\EE[\gamma_1] &=& g_1^\mmse\left(\alpha,\beta-\frac{1}{N}\right)\\
& = & g_1^\mmse\left( \alpha,\beta \right) - \frac{1}{N}
\frac{\partial}{\partial \beta}  g_1^\mmse\left( \alpha,\beta
\right) + O\left(N^{-2}\right),
\end{eqnarray}
where
\begin{eqnarray} \label{eq:gam1_bar_mmse}
\frac{\partial}{\partial \beta} g_1^\mmse\left( \alpha,\beta
\right) &=& - \frac{\alpha}{2} \left[ 1
+\frac{\alpha(1-\beta)-1}{\sqrt{(\alpha (1-\beta)-1)^2+4\alpha}}
\right].
\end{eqnarray}
For later use, we define also the following asymptotic moments
\begin{eqnarray}\label{g_k_def}
    g_m^\mmse(\alpha,\beta) & \triangleq & \lim_{N \rightarrow \infty} \alpha^m  \EE \left[ \frac{1}{N} {\rm Tr} \left (
    \left[\Id + \alpha  \Hm_1 \Hm_1^\herm \right]^{-m} \right )
    \right],
\end{eqnarray}
which can be obtained by repeatedly differentiating
$g_1^\mmse(\alpha,\beta)$ with respect to $\alpha$ using the
recursive relation
\begin{eqnarray}\label{g_k_rec}
    g_{m+1}^\mmse(\alpha,\beta) & = & \frac{\alpha^2}{m} \frac{\partial}{\partial \alpha} g_m^\mmse(\alpha,\beta), \;\;\;\; m \geq 1.
\end{eqnarray}
Thus we have
\begin{eqnarray}\label{g_2_def}
    g_2^\mmse(\alpha,\beta) & = & \alpha^2 \frac{\partial}{\partial \alpha} g_1^\mmse(\alpha,\beta), \\ \label{g_3_def}
    g_3^\mmse(\alpha,\beta) & = & \frac{\alpha^3}{2} \left(\alpha \frac{\partial^2}{\partial \alpha^2} g_1^\mmse(\alpha,\beta)
    + 2 \frac{\partial}{\partial \alpha} g_1^\mmse(\alpha,\beta) \right).
\end{eqnarray}
For large SNR, i.e. $\alpha=\rho/\beta\gg 1$ and $\beta<1$,
$g_1^\mmse(\alpha,\beta)$ is approximately $\alpha(1-\beta)$. This result
indicates that only the $\approx N(1-\beta)$ zero eigenvalues of the
matrix $\Hm_1 \Hm_1^\herm$ contribute to the SINR for large
$\alpha$. Similarly, $g_m^\mmse(\alpha,\beta) \approx (1-\beta) \alpha^m$
for large $\rho$ and $\beta<1$, while for $\beta=1$,
$g_m^\mmse(\alpha,\beta) \approx k_m \rho^{m-1/2}$, where the constant
$k_m$ satisfies $k_{m+1} = \prod_{j=1}^m\left(1-1/2j\right)$.

Next we calculate the matrix of the joint cumulants of order 2 with elements
\[ \Sigma_{i,j}^\mmse = \EE_c\left[\gamma_i;\gamma_j\right] \equiv
\EE\left[\gamma_i\gamma_j\right] - \EE\left[\gamma_i\right]
\EE\left[\gamma_j\right]. \]
Given the symmetry, all diagonal
elements ($i = j$) are equal, and so are all off-diagonal ones ($i
\neq j$). Therefore, it is sufficient to compute
$\Sigma_{1,1}^\mmse$ and $\Sigma_{1,2}^\mmse$.

We start with $\EE_c\left[\gamma_1;\gamma_1\right]$. For
convenience, we define $\Bm_1 \triangleq \left( \Id + \alpha\Hm_1
\Hm_1^\herm \right)^{-1}$, and let $(\Bm_1)_{ij}$ denote the
$(i,j)^\text{th}$ element of $\Bm_1$ and $h_{1i}$ denote the
$i^\text{th}$ element of $\hv_1$. Then, a direct application of
(\ref{eq:sir_def}) and (\ref{eq:novikov_thm_2}) yields
\begin{eqnarray}\label{eq:autocorr_gamma}
  \Sigma^\mmse_{1,1} & = & \EE_c[\gamma_1;\gamma_1] \nonumber \\
  &=& \alpha^2 \sum_{a,b,c,d} \EE\left[(\Bm_1)_{ab} (\Bm_1)_{cd} \left(h^*_{1a} h_{1b} h^*_{1c} h_{1d} - \frac{\delta_{a,b}}{N}
  \frac{\delta_{c,d}}{N} \right)
  \right] \nonumber \\
 &=& \alpha^2 \sum_{a,b,c,d} \EE\left[(\Bm_1)_{ab}(\Bm_1)_{cd}\right] \EE_c\left[h^*_{1a} h_{1b}; h^*_{1c} h_{1d} \right]%
  \nonumber \\
 &= & \alpha^2 \EE\left[ \frac{1}{N^2} {\rm Tr} (\Bm_1^2) \right] \nonumber \\
  &\rightarrow & \frac{g^\mmse_2 \left(\alpha,\beta-\frac{1}{N}\right) }{N}  =  \frac{v_d^\mmse}{M} + O(1/N^2)
\end{eqnarray}
where $v^\mmse_d = \beta g_2^\mmse \left( \alpha,\beta-\frac{1}{N}
\right)$. We see that the leading correction in the autocorrelation
is non-vanishing only due to the random character of the vector
$\hv_1$ \cite{Tse2000_MMSEFluctuations}.

We now turn to the more complicated computation of $\Sigma^\mmse_{1,2}$
to leading order in $N$. To simplify notation, we define the
matrices $\Bm_{i} = \left ( \Id + \alpha\Hm_{i}\Hm_{i}^\herm \right )^{-1}$, for
$i=1,2$, as before, and $\Bm_{12} = \left (\Id + \alpha
\Hm_{12}\Hm_{12}^\herm \right )^{-1}$ where $\Hm_{12}$ is obtained by striking
out from $\Hm$ both columns $\hv_1$ and $\hv_2$. Therefore,
\begin{eqnarray}\label{eq:gamma_12}
  \gamma_1 &=& \alpha\hv_1^\herm \Bm_1 \hv_1 \nonumber \\
  \gamma_2 &=& \alpha \hv_2^\herm  \Bm_2 \hv_2
\end{eqnarray}

Using the same notation as before, we rewrite the cumulant moment of $\gamma_1$,
$\gamma_2$ as
\begin{eqnarray}\label{eq:gamma12_corr}
\EE_c[\gamma_1;\gamma_2] = \alpha^2 \sum_{abcd}
\EE_c\left[h_{1a}^* \left(\Bm_1\right)_{ab} h_{1b} ; h_{2c}^*
\left(\Bm_2\right)_{cd} h_{2d}\right]
\end{eqnarray}
In the following we will make extensive use of the following matrix
identities, obtained by applying the Sherman-Morrison matrix
inversion lemma,
\begin{eqnarray}\label{eq:ident_expan}
\Bm_2 & = & \Bm_{12} - \Bm_{12}\hv_1 \hv_1^\herm \Bm_{12}
\frac{\alpha}{1+\alpha\hv_1^\herm \Bm_{12} \hv_1} \nonumber \\
\Bm_1 & = & \Bm_{12} - \Bm_{12}\hv_2 \hv_2^\herm \Bm_{12}
\frac{\alpha}{1+\alpha\hv_2^\herm \Bm_{12} \hv_2}
\end{eqnarray}
We will now use Novikov's theorem (Theorem~\ref{thm:Novikov} in
Appendix~\ref{app:novikov_thm}) to successively average over the
variables $\hv_1$ and $\hv_2$. For example, considering the general
term for indices $(a,b,c,d)$ in (\ref{eq:gamma12_corr}) we write
\begin{eqnarray}
& & \EE_c\left[h_{1a}^* \left(\Bm_1\right)_{ab} h_{1b} ; h_{2c}^* \left(\Bm_2\right)_{cd} h_{2d}\right] = \nonumber \\
& = & \EE \left [ h_{1a}^* \left(\Bm_1\right)_{ab} h_{1b} h_{2c}^* \left(\Bm_2\right)_{cd} h_{2d}\right] -
\EE \left [ h_{1a}^* \left(\Bm_1\right)_{ab} h_{1b} \right ] \EE \left [ h_{2c}^* \left(\Bm_2\right)_{cd} h_{2d}\right] \nonumber \\
& = & \frac{1}{N} \EE \left[ \frac{\partial}{\partial h_{1a}} \left(
(\Bm_1)_{ab} (\Bm_2)_{cd} h_{1b} h_{2c}^* h_{2d} \right) \right] -
\frac{1}{N}
\EE \left[ \frac{\partial }{\partial h_{1a}} \left( (\Bm_1)_{ab} h_{1b} \right) \right] \EE \left[ h_{2c}^* (\Bm_2)_{cd} h_{2d} \right] \label{novikov-appl} \\
& = & \frac{1}{N} \EE \left[ \frac{\partial h_{1b}}{\partial h_{1a}}
(\Bm_1)_{ab} (\Bm_2)_{cd} h_{2c}^* h_{2d} \right] +
\frac{1}{N} \EE \left[ \frac{\partial (\Bm_2)_{cd}}{\partial h_{1a}} (\Bm_1)_{ab} h_{1b} h_{2c}^* h_{2d} \right] \nonumber \\
& & - \frac{1}{N}  \EE \left[ \frac{\partial h_{1b}}{\partial h_{1a}} (\Bm_1)_{ab}  \right] \EE \left[ h_{2c}^* (\Bm_2)_{cd} h_{2d} \right]
\label{eq:cumulant_compute}
\end{eqnarray}
where in (\ref{novikov-appl}) we have applied Novikov's theorem
formally replacing $h_{1a}^*$ with $\frac{1}{N}
\frac{\partial}{\partial h_{1a}}$ inside the expectations. We remind
the reader that, as explained in Section \ref{app:novikov_thm}, in
the above manipulations we treat the complex variables $h_{ka}$ and
$h_{nb}^*$ as distinct and independent for all $k,n,a,b$, such that
partial derivatives are performed individually with respect to these
variables. In order to compute $\frac{\partial
(\Bm_2)_{cd}}{\partial h_{1a}}$ we use the matrix inversion lemma
(\ref{eq:ident_expan}) for $\Bm_2$. After some algebra, we obtain
\begin{eqnarray}\label{eq:gamma12_corr2_1}
\EE_c[\gamma_1;\gamma_2] &=& \frac{\alpha^2}{N} \EE_c\left[ {\rm
Tr}\left(
\Bm_1\right) ; \hv_{2}^\herm \Bm_2 \hv_{2}\right] \\ \label{eq:gamma12_corr2_2} %
&-&\frac{\alpha^3}{N} \EE\left[ \frac{\hv_2^\herm \Bm_{12}\Bm_1
\hv_1 \hv_{1}^\herm \Bm_{12} \hv_{2}}{1+\alpha\hv_{1}^\herm
\Bm_{12} \hv_{1}}\right]
\\ \label{eq:gamma12_corr2_3} %
&+&\frac{\alpha^4}{N} \EE\left[ \frac{\hv_1^\herm \Bm_{12}\Bm_1
\hv_1 \hv_{2}^\herm \Bm_{12} \hv_{1}\hv_{1}^\herm \Bm_{12} \hv_{2}
}{\left(1+\alpha\hv_{1}^\herm \Bm_{12} \hv_{1}\right)^2}\right].
\end{eqnarray}
The term in \eqref{eq:gamma12_corr2_1} results by summing over all
indices the first two terms in \eqref{eq:cumulant_compute}, and the
terms in \eqref{eq:gamma12_corr2_2} and \eqref{eq:gamma12_corr2_3}
result by summing the last term in  \eqref{eq:cumulant_compute}
after applying the partial derivative with respect to the elements
of $\hv_1$ appearing in the numerator and the denominator of the
matrix inversion lemma expansion of $\Bm_2$. It is important to
notice that the order of magnitude of the first term is $O(1)$,
while the last two terms are $O(1/N)$. The reason is that the last
two terms are the result of applying the partial derivative in
$h_{1a}$ to $\Bm_2$, where the term that depends on $\hv_1$ is
scaled by a factor $O(1/N)$ compared to the remaining matrix.

We proceed now by applying Novikov's theorem to the random
variables $h_{2a}^*$ appearing in the numerator of (\ref{eq:gamma12_corr2_2}) and (\ref{eq:gamma12_corr2_3})
and exchanging the corresponding expectation with a derivative
$\partial/\partial h_{2a}$. However, with some hindsight we only apply the derivative
to $\hv_2$ and not to $\Bm_1$, which would give a subleading term in $1/N$.
Therefore, to leading order in $1/N$, we have
\begin{eqnarray}\label{eq:gamma12_corr2_4}
(\ref{eq:gamma12_corr2_2}) &\approx& - \frac{\alpha^3}{N^2}
\EE\left[ \frac{\hv_1^\herm \Bm_{12}^2\Bm_1
\hv_1}{1+\alpha\hv_{1}^\herm \Bm_{12} \hv_{1}}\right] \approx -
\frac{\alpha^3}{N^2} \frac{\frac{{\rm
Tr}\left(\Bm_{12}^3\right)}{N}}{1+\frac{\alpha}{N}{\rm Tr}\left(
\Bm_{12} \right)}\approx - \frac{1}{N^2} \frac{g^\mmse_3\left(\alpha,
\beta-\frac{2}{N} \right)}{1+g^\mmse_1\left(\alpha, \beta-\frac{2}{N}
\right)}
\\ \label{eq:gamma12_corr2_5} %
(\ref{eq:gamma12_corr2_3}) &\approx&  \frac{\alpha^4}{N^2}
\EE\left[ \frac{\hv_1^\herm \Bm_{12}\Bm_1 \hv_1 \hv_1^\herm
\Bm_{12}^2 \hv_1}{(1+\alpha\hv_{1}^\herm \Bm_{12}
\hv_{1})^2}\right] \approx \frac{\alpha^4}{N^2}
\frac{\left(\frac{{\rm
Tr}\left(\Bm_{12}^2\right)}{N}\right)^2}{\left(1+\frac{\alpha}{N}{\rm
Tr}\left( \Bm_{12} \right)\right)^2} %
\approx  \frac{1}{N^2} \frac{g^\mmse_2\left(\alpha, \beta-\frac{2}{N}
\right)^2}{\left( 1+g^\mmse_1\left(\alpha, \beta-\frac{2}{N} \right)
\right)^2}
\end{eqnarray}
where the approximation sign $\approx$ means to leading order in
$1/N$. The second expression in each line occurred by averaging over
$\hv_1$ to leading order, i.e. only on the numerator. In the last
equation in each line we used the fact that
$\frac{1}{N} {\rm Tr} ( \Bm_{12})  \approx g^\mmse_1\left(\alpha, \beta-\frac{2}{N} \right)$.

Next we may go back to (\ref{eq:gamma12_corr2_1}) and expand $\Bm_2$
using (\ref{eq:ident_expan}). After applying exactly the same
methods as above we arrive at the following expression
\begin{eqnarray}\label{eq:gamma12_corr2_6}
(\ref{eq:gamma12_corr2_1}) &\approx& \frac{\alpha^2}{N^2}
\EE_c\left[ {\rm Tr}\left( \Bm_{12}\right) ; {\rm Tr}\left(
\Bm_{12}\right)\right] + \frac{2}{N^2} \frac{g^\mmse_2\left(\alpha,
\beta-\frac{2}{N} \right)^2}{\left(1+g^\mmse_1\left(\alpha,
\beta-\frac{2}{N} \right)\right)^2} - \frac{1}{N^2}
\frac{g^\mmse_3\left(\alpha, \beta-\frac{2}{N} \right)}{1+g^\mmse_1\left(\alpha,
\beta-\frac{2}{N} \right)}.
\end{eqnarray}
We collect all terms and use (\ref{eq:cum_moment2_eigs}) to reach
the final result,
\begin{eqnarray}\label{gamma_xcorr1_6}
  \Sigma_{1,2}^\mmse & = & \EE_c[\gamma_1;\gamma_2] \nonumber \\
    &\approx &\frac{1}{N^2}\frac{\left(\beta-\frac{2}{N}\right) \alpha^4}{\left(1+2\alpha\left(1+\beta-\frac{2}{N}\right)+\alpha^2\left(1-\beta+\frac{2}{N}\right)^2\right)^2}
    \\ \nonumber
  &+& \frac{1}{N^2}\left(
   \frac{3 g^\mmse_2\left(\alpha,
\beta-\frac{2}{N} \right)^2}{\left(1+g^\mmse_1\left(\alpha,
\beta-\frac{2}{N} \right) \right)^2}
    - \frac{2g^\mmse_3\left(\alpha,
\beta-\frac{2}{N} \right)}{1+g^\mmse_1\left(\alpha, \beta-\frac{2}{N}
\right)}\right) \\ \nonumber
& = & \frac{v_{od}^\mmse}{M^2} + O(1/N^3),
\end{eqnarray}
where we let
\begin{eqnarray} \label{eq:vod_mmse}
v_{od}^\mmse & = & \beta^2\left(   \frac{3 g^\mmse_2\left(\alpha,
\beta-\frac{2}{N}\right)^2}{\left(1+g^\mmse_1\left(\alpha,
\beta-\frac{2}{N}\right) \right)^2}
    - \frac{2g^\mmse_3\left(\alpha, \beta-\frac{2}{N}\right)}{1+g^\mmse_1\left(\alpha,
    \beta-\frac{2}{N}\right)} \nonumber \right.\\
&& \left.
+\frac{\left(\beta-\frac{2}{N}\right)\alpha^4}{\left(1+2\alpha\left(1+\beta-\frac{2}{N}\right)+\alpha^2\left(1-\beta+\frac{2}{N}\right)^2\right)^2}\right).
\end{eqnarray}
For large $\alpha$, $v_{od}^\mmse \approx \rho^2$ when $\beta<1$,
and $v_{od}^\mmse \approx \rho^2/16$ when $\beta=1$.

We collect the results of (\ref{gamma_xcorr1_6}) and
(\ref{eq:autocorr_gamma}) by writing the correlation matrix for the SINRs $\{\gamma_k\}$ to leading order as:
\begin{eqnarray}\label{gamma_corr_matrix}
\Sigma_{i,j}^\mmse = \delta_{i,j} \frac{v_d^\mmse}{M} +
(1-\delta_{i,j}) \frac{v_{od}^\mmse}{M^2}
\end{eqnarray}
It is worth pointing out that despite the fact that the
off-diagonal elements are much smaller compared to the diagonal
ones, they all contribute to the eigenvalues of $\Sigmam$.
In fact, these can be computed in closed form and are given by
\[ \lambda_1(\Sigmam^\mmse) = \frac{v_d^\mmse}{M} + (M-1) \frac{v_{od}^\mmse}{M^2} \approx \frac{v_d^\mmse+v_{od}^\mmse}{M} \]
and
\[ \lambda_k(\Sigmam^\mmse) =  \frac{v_d^\mmse}{M} - \frac{v_{od}^\mmse}{M^2} \approx \frac{v_d^\mmse}{M} \]
for all $k = 2,\ldots,M$.

\subsubsection{Cumulant moments for the ZF receiver}\label{sec:cum_moments12_ZF}

The corresponding results for the ZF receiver can be derived
directly from the previous section by observing that the SINR for
the $k$-th channel of the ZF receiver, given by  (\ref{sinr-zf}), can be deduced from the corresponding
expression (\ref{sinr-mmse}) for the MMSE receiver in the limit of infinite $\alpha$, i.e.
\begin{eqnarray}\label{eq:gammaZF_MMSE_derivation}
\gamma_k^{\zf} =
\frac{\alpha}{\left[\left(\Hm^\herm\Hm\right)^{-1}\right]_{kk}}
&=& \alpha \lim_{\alpha_0\rightarrow\infty}
\frac{\gamma_k^\mmse(\alpha_0)}{\alpha_0} \\ \nonumber %
&=& \alpha \lim_{\alpha_0\rightarrow\infty} \left\{\alpha_0\left[\left(\Id
+\alpha_0\Hm^\herm\Hm\right)^{-1}\right]_{kk}\right\}^{-1}
\end{eqnarray}
A subtle point needs to be
stressed here:
the results for the ZF receiver cannot be obtained simply
as the ``limit for high SNR'' of the results for the MMSE receiver.
Rather, we have to distinguish between the channel SNR (contained in the parameter $\alpha$) and the SNR
parameter in the linear receiver matrix expression (indicated by
$\alpha_0$ above) that we let to infinity in order to obtain the ZF
results. It can be shown that for $\beta<1$ the analysis of the
previous section involving the matrices $\Bm_1, \Bm_2$ and
$\Bm_{12}$ can be carried out in this limiting case. In addition, as
seen in Appendix \ref{app:eig_flucts}, the condition for the
validity of the manipulation of the first term of
(\ref{gamma_xcorr1_6}) is that $f(x)=(1+\alpha x)^{-1}$ is a smooth
function of $x$ in the region of support of the eigenvalue spectrum.
This is not true in the vicinity of $x = 0$ for arbitrarily large
$\alpha$, specifically when $\alpha = O(N)$. Thus when $\beta=1$, in
which case the asymptotic eigenvalue spectrum includes $x=0$, the
above approximation is not valid. As a result, this method breaks
down at $\beta=1$.

From (\ref{eq:mean_sir__plus_correction}) we get the mean SINR for the ZF receiver\footnote{For simplicity we neglect the subleading
terms in the following equalities, i.e., we omit $O(1/N^2)$ in (\ref{eq:gammaZF_mean}) and
$O(1/N^3)$ in (\ref{eq:gammaZF_sigma11}), (\ref{eq:gammaZF_sigma12}), respectively.}
\begin{equation}\label{eq:gammaZF_mean}
    \EE[\gamma_1^\zf] = \left\{\begin{array}{c c} \alpha(1-\beta + 1/N)  & \beta<1 \\
                              0  & \beta=1 \\
                                \end{array} \right.
\end{equation}
in agreement with \cite{Tse2000_MMSEFluctuations}. Similarly, the
second order moments can be obtained from
\begin{eqnarray}\label{eq:gammaZF_2nd_moment_derivation}
\EE_c\left[\gamma_i^\zf;\gamma_j^\zf\right] = \alpha^2
\lim_{\alpha_0\rightarrow\infty}
\frac{\EE_c\left[\gamma_i^\mmse(\alpha_0);\gamma_j^\mmse(\alpha_0)\right]}{\alpha_0^2}.
\end{eqnarray}
Thus we get
\begin{equation}\label{eq:gammaZF_sigma11}
    \Sigma_{11}^\zf = \frac{v_d^\zf}{M} = \left\{\begin{array}{c c} \frac{\alpha^2\beta(1-\beta + 1/N)}{M} & \beta<1 \\
                              0  & \beta=1 \\
                            \end{array} \right.
\end{equation}
and
\begin{equation}\label{eq:gammaZF_sigma12}
    \Sigma_{12}^\zf = \frac{v_{od}^\zf}{M^2} = \left\{\begin{array}{c c} \frac{\alpha^2\beta^2}{M^2} & \beta<1 \\
                              \frac{\rho^2}{16M^2}  & \beta=1 \\
                            \end{array} \right. .
\end{equation}
While in the case of $\beta<1$ the covariance
matrix $\Sigmam$ is well-defined and positive-definite with
eigenvalues
\[ \lambda_1(\Sigmam^\zf) = \frac{v_d^\zf}{M} + (M-1) \frac{v_{od}^\zf}{M^2} \approx \frac{\alpha^2 \beta^2}{\beta M} \]
and
\[ \lambda_k(\Sigmam^\zf) =  \frac{v_d^\zf}{M} - \frac{v_{od}^\zf}{M^2} \approx \frac{\alpha^2\beta^2(1-\beta)}{\beta M} \]
for all $k = 2,\ldots,M$,
the case $\beta=1$ is problematic.
Specifically, it results in (narrowly) negative eigenvalues for
$\Sigmam$, thereby invalidating the Gaussian approximation for the
$\gamma_i$'s and as a result the further treatment of the mutual
information as a Gaussian variable. The case $\beta = 1$ for the ZF
receiver is therefore excluded in the subsequent Gaussian
approximation of the mutual information.

\subsection{Gaussian approximation and outage probability}

In this section we use the previous results together with the
asymptotic Gaussianity that follows from the fact that higher-order
moments are vanishing (see Appendix \ref{app:asympt_cum}) to give an
explicit Gaussian approximation for the outage probability of linear
MMSE and ZF receivers with coding across the antennas in the regime
of fixed SNR and large number of antennas.

We start with the mean $\EE[I_N]$. Due to the symmetry with
respect to the terms $\gamma_k$, we have
\begin{eqnarray}\label{eq:MI_MMSE_mean}
  {\cal C}_1 = \EE[I_N] &=& M \EE[ \log(1+\gamma_1)]  =
  M \EE[ \log(1+\EE[\gamma_1] + \gamma_1 - \EE[\gamma_1])] \\ \nonumber %
  &=& M\log(1+\EE[\gamma_1]) - M \sum_{n=1}^\infty \frac{(-1)^n}{n}
  \frac{\EE\left[\left(\gamma_1-\EE[\gamma_1]\right)^n\right]}{\left(1+\EE[\gamma_1]\right)^n}
\end{eqnarray}
In the above expansion, all terms $n>2$ involve cumulants of
$\gamma_1$ higher than 2, thus can be neglected. For the MMSE
receiver, this yields
\begin{eqnarray} \label{eq:C_1_mmse}
{\cal C}_1^\mmse = \EE[I_N] &=& M \log \left( 1 + g_1^\mmse\left(
\alpha,\beta-\frac{1}{N} \right) \right) - \frac{v_d^\mmse}{2
(1+g_1^\mmse\left( \alpha,\beta-\frac{1}{N} \right))^2} + o(1)\nonumber\\
&\approx& M \log \left( 1 + g_1^\mmse(\alpha,\beta) - \frac{1}{N}
\frac{\partial}{\partial \beta} g_1^\mmse(\alpha,\beta) \right) -
\frac{v_d^\mmse}{2 (1+g_1^\mmse\left( \alpha,\beta \right))^2} \nonumber \\
&\approx& M \log ( 1 + g^\mmse_1(\alpha,\beta)) - \beta
\frac{\frac{\partial}{\partial
\beta} g_1^\mmse(\alpha,\beta)}{1+g_1^\mmse(\alpha,\beta)}
- \frac{v_d^\mmse}{2 (1+g_1^\mmse\left( \alpha,\beta \right))^2},
\end{eqnarray}
%
where $g_1^\mmse(\alpha,\beta)$ and $v_d^\mmse$ are given by
\eqref{eq:ave_SIR} and (\ref{eq:autocorr_gamma}) respectively.

For the ZF receiver the mean is given  by
\begin{eqnarray}\label{eq:MI_ZF_mean1}
{\cal C}_1^\zf = M \log(1+\alpha(1-\beta)) + \frac{\alpha \beta
\left(1+\frac{\alpha(1-\beta)}{2}\right)}{(1+\alpha(1-\beta))^2} +
o(1).
\end{eqnarray}

Similarly, we can calculate the variance of the mutual information as follows
\begin{eqnarray}\label{eq:Var_MI_MMSE}
  {\cal C}_2  &=& M \EE_c[\log(1+\gamma_1); \log(1+\gamma_1)] +
  M(M-1)\EE_c[\log(1+\gamma_1); \log(1+\gamma_2)] \nonumber \\
                &=& M \sum_{m,n=1}^\infty \frac{(-1)^{m+n}}{mn}
                \frac{\EE_c[\left(\gamma_1 -\EE[\gamma_1]\right)^m; \left(\gamma_1-\EE[\gamma_1]\right)^n]}
                {\left(1+\EE[\gamma_1]\right)^{n+m}} \nonumber \\
                &+& M(M-1) \sum_{m,n=1}^\infty \frac{(-1)^{m+n}}{mn}
                \frac{\EE_c[\left(\gamma_1-\EE[\gamma_1]\right)^m; \left(\gamma_2-\EE[\gamma_2]\right)^n]}
                {\left(1+\EE[\gamma_1]\right)^{n+m}}.
\end{eqnarray}
As we see above, the first terms in both the above summations give
the leading order of the variance of the mutual information, which
we denoted in (\ref{eq:asympt_C_2}) by $\sigma^2$. The variance of
the MMSE mutual information is given by
\begin{eqnarray}\label{eq:MI_MMSE_sigma1}
{\cal C}_2^\mmse &=& \frac{v_d^\mmse +
v_{od}^\mmse}{\left(1+g_1^\mmse\left(\alpha,\beta-\frac{1}{N}\right)\right)^2}
+ o(1),
\end{eqnarray}
where $v_{od}^\mmse$ is given by \eqref{eq:vod_mmse}. The
corresponding variance for the ZF receiver is
\begin{eqnarray}\label{eq:MI_ZF_sigma1}
  {\cal C}_2^\zf &=&
  \frac{\beta \alpha^2 (1 + 1/N)}{\left(1+\alpha(1-\beta + 1/N)\right)^2} + o(1)
\end{eqnarray}
for $\beta<1$. As mentioned above, the case $\beta=1$
does not result in a well-behaved jointly Gaussian
behavior of the $\gamma_k$'s, and therefore the Gaussian
approximation of the mutual information cannot be derived with
this approach.

As anticipated at the beginning of this section, from (\ref{eq:C_1_mmse}) and (\ref{eq:MI_ZF_mean1})
we see that the mean mutual information is expressed in the form $m_1 = Mc_{10} + c_{11}$, where
the coefficient $c_{10}$ was found in previous works (e.g., \cite{verdu-shamai99}),
considering the limit $N \rightarrow \infty$ of the  normalized mutual information per transmit antenna,
and the term $c_{11}$ is a correction term that captures the correlation between the SINRs $\gamma_k$.

Under this Gaussian approximation, we can easily evaluate the outage probability
for fixed SNR, $\beta$ and number of antennas $M$ as follows:
\begin{equation}\label{eq:gaussian_approx_mmse_MI}
P_{\rm out}^\text{lin}(R,\rho) \approx Q \left ( \frac{R - {\cal C}_1}{\sqrt{{\cal C}_2}} \right )
\end{equation}
where $Q(x) = \int_x^\infty \frac{1}{\sqrt{2\pi}} e^{-t^2/2} dt$ is the Gaussian tail function.

We conclude this section with a discussion on the range of validity
of the Gaussian approximation. For the MMSE receiver, in the large
$\rho$ limit we have that $\Cc_2^\mmse = O(\rho)$ for $\beta=1$,
while ${\cal C}_2^\mmse = O(1)$ for $\beta < 1$. This fast increase
of the variance of the distribution for $\beta=1$ and $\rho \gg 1$
is a spurious result in this approximation, due to the neglected
terms which are negligible for fixed $\rho$ and increasingly large
$N$, but become important for fixed $N$ and large $\rho$.

The behavior of the ZF receiver for $\beta=1$, when the
jointly Gaussian behavior of the $\gamma_k$'s breaks down,
exacerbates the above large $\rho$ behavior of the MMSE receiver.
In fact, as was discussed before, the ZF case is in some sense the
``infinite $\rho$ limit'' of the MMSE case. Therefore the problematic
situation appearing in the ZF receiver for $\beta=1$ has the same
roots as the problems faced in the large (but finite) $\rho$ limit
when $N$ is also finite but not large enough.

\subsection{Simulations and comparisons}

In this section we first validate the asymptotic analysis by
comparing the asymptotic approximation for $\Cc_1$ and $\Cc_2$ with
the exact moments obtained by finite-dimensional Monte Carlo
simulation. We then compare the Gaussian approximation to the outage
probability with finite-dimensional Monte Carlo simulation.

For the sake of comparison, we also consider the outage probability
of the optimal receiver, given by the log-det cdf $P( I_N^{\sf opt}
\leq R)$, where $I_N^{\sf opt} = \log \det (\Id + \frac{\rho}{M}
\Hm\Hm^\herm)$. As said in the Introduction, the asymptotic
Gaussianity of the log-det mutual information is well-known and
holds under very general models of channel correlation across the
antennas (not considered in this work). For completeness, we recall
its expressions under the i.i.d. channel coefficient assumptions and
in the notation of this paper. We have
\begin{equation}
\frac{I_N^{\rm opt} - M \mu }{\nu} \; \stackrel{d}{\rightarrow} \Nc(0,1)
\end{equation}
where
\begin{equation} \label{mu}
\mu = \log(1 + g^\mmse_1(\alpha,\beta)) + \frac{1}{\beta} \log(1 -
\alpha(1 - \beta) + g^\mmse_1(\alpha,\beta)) +
\frac{g^\mmse_1(\alpha,\beta)}{\alpha\beta} - \frac{1}{\beta},
\end{equation}
and
\begin{equation} \label{sigma2}
\nu^2 = - \log \left (1 - \frac{1}{\beta} \left (1 -
\frac{g^\mmse_1(\alpha,\beta)}{\alpha} \right )^2 \right ).
\end{equation}
The first term in (\ref{mu}) coincides with the coefficient of $M$ in the first
term of \eqref{eq:C_1_mmse}, i.e., it is the asymptotic capacity per
antenna of the linear MMSE receiver. The additional terms in
(\ref{mu}) represent the so-called ``non-linear'' gain of the
optimal versus linear MMSE receiver, as discovered in
\cite{verdu-shamai01}.

It is worth pointing out that for large SNR, the asymptotic mean capacity is given by
\begin{eqnarray}\label{eq:asympt_optimal_mi}
\EE[ I_N^{\rm opt}]  \approx  M \log \rho
\end{eqnarray}
while the variance has the following behavior
\begin{equation}\label{eq:asympt_optimal_var_mi}
    \nu^2 \approx \left\{\begin{array}{c c} -\log (1-\beta) & \beta<1 \\
                              \frac{1}{2} \log \rho  & \beta=1 \\
                            \end{array} \right.
\end{equation}
Using the Gaussian approximation
\[ P_{\rm out}(R, \rho) \approx Q\left ( \frac{R - M\mu}{\nu} \right ) \approx \exp \left ( - \frac{(R - M\mu)^2}{2\nu^2} \right ) \]
with $R = r\log\rho$,  in the large $\rho$ limit we find
\begin{eqnarray}\label{eq:asympt_P_out}
    \left\{\begin{array}{c c c} - \frac{\log P_{out}(R,\rho)}{\log \rho} \approx \left(M - r \right)^2 \frac{\log \rho }{2|\log(1-\beta)|} && \beta<1 \\
                              - \frac{\log P_{out}(R, \rho)}{\log \rho} \approx \left(M - r \right)^2       && \beta=1 \\
                            \end{array} \right.
\end{eqnarray}
For $\beta = 1$,  the Gaussian approximation yields
an outage probability exponent equal to $(r - M)^2$ for $r \in
[0,M]$, that closely approximates the exact exponent $d^*(r)$
\cite{ZheTse}. However, for $\beta < 1$ the Gaussian approximation
yields a completely inaccurate behavior. In fact, in this case the
exponent obtained through (\ref{eq:asympt_P_out}) would be infinite,
while we know from \cite{ZheTse} that $d^*(r) \leq MN$. The reason
for this spectacular failure of the Gaussian approximation is that,
in the large-$N$ approximation, it is implicitly assumed that for
$\beta \neq 1$ the eigenvalue distribution at very small eigenvalues
is zero, as described by the Marcenko-Pastur law
\cite{Tulino2004_RMTInfoTheoryReview}. Instead, for large $\rho$ the
eigenvalues that dominate the outage probability are exactly the
very small ones, of the order of $1/\rho$, i.e., exactly the ones
that the Gaussian approximation neglects.

\begin{figure}[htbp]
\begin{center}
\hspace*{-0.62in}
\includegraphics[width=20.5cm]{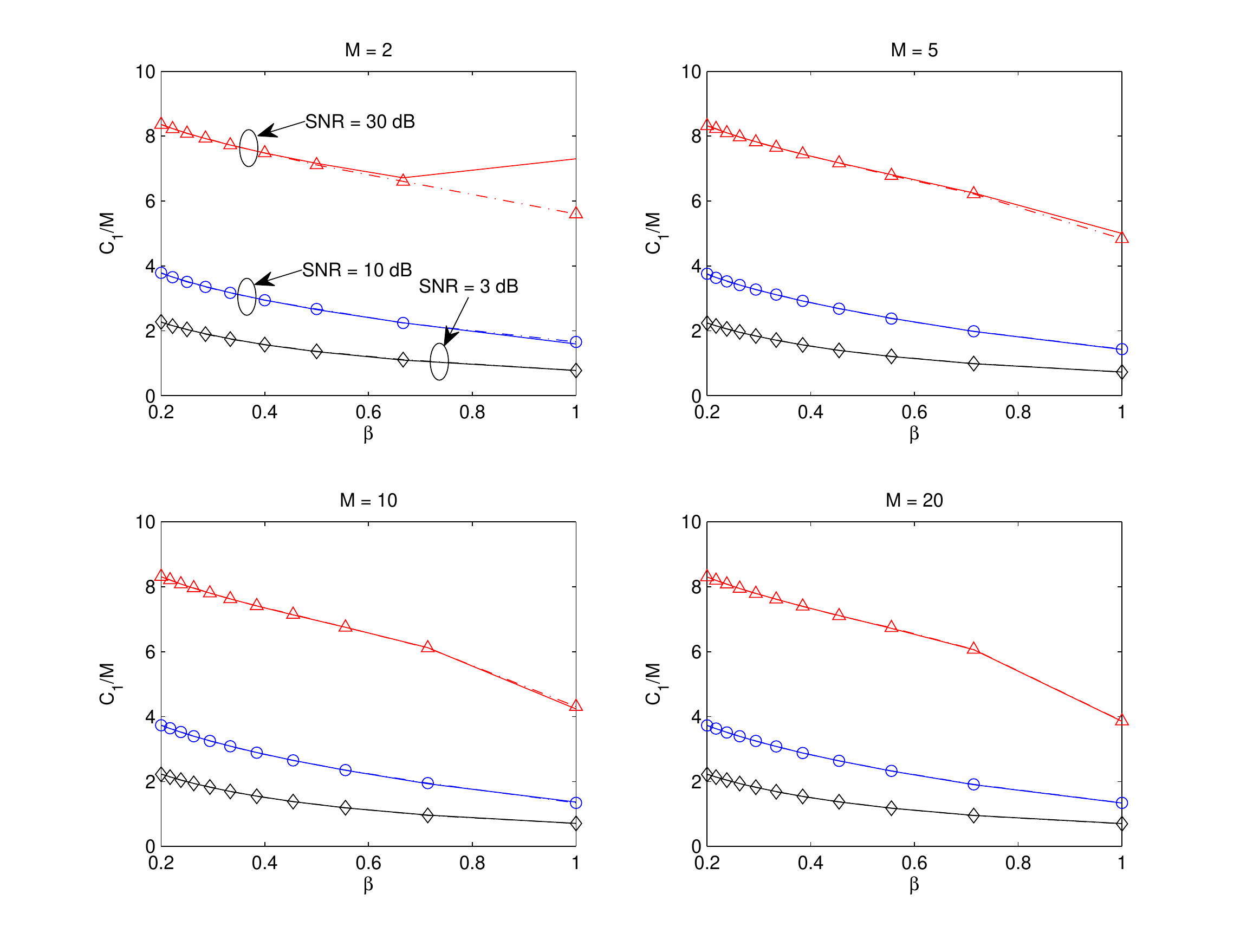}
\caption{Mean of the MMSE mutual information per antenna (${\cal
C}_1^\mmse/M$) as a function of $\beta$, for $M = 2,5,10$ and $20$.
The solid lines are analytical results, and the corresponding
dash-dot lines are empirical results obtained from Monte Carlo
simulation. Diamonds denote 3 dB, circles 10 dB and triangles 30
dB.} \label{fig:MMSE_mean}
\end{center}
\end{figure}

\begin{figure}[htbp]
\begin{center}
\hspace*{-0.65in}
\includegraphics[width=21cm]{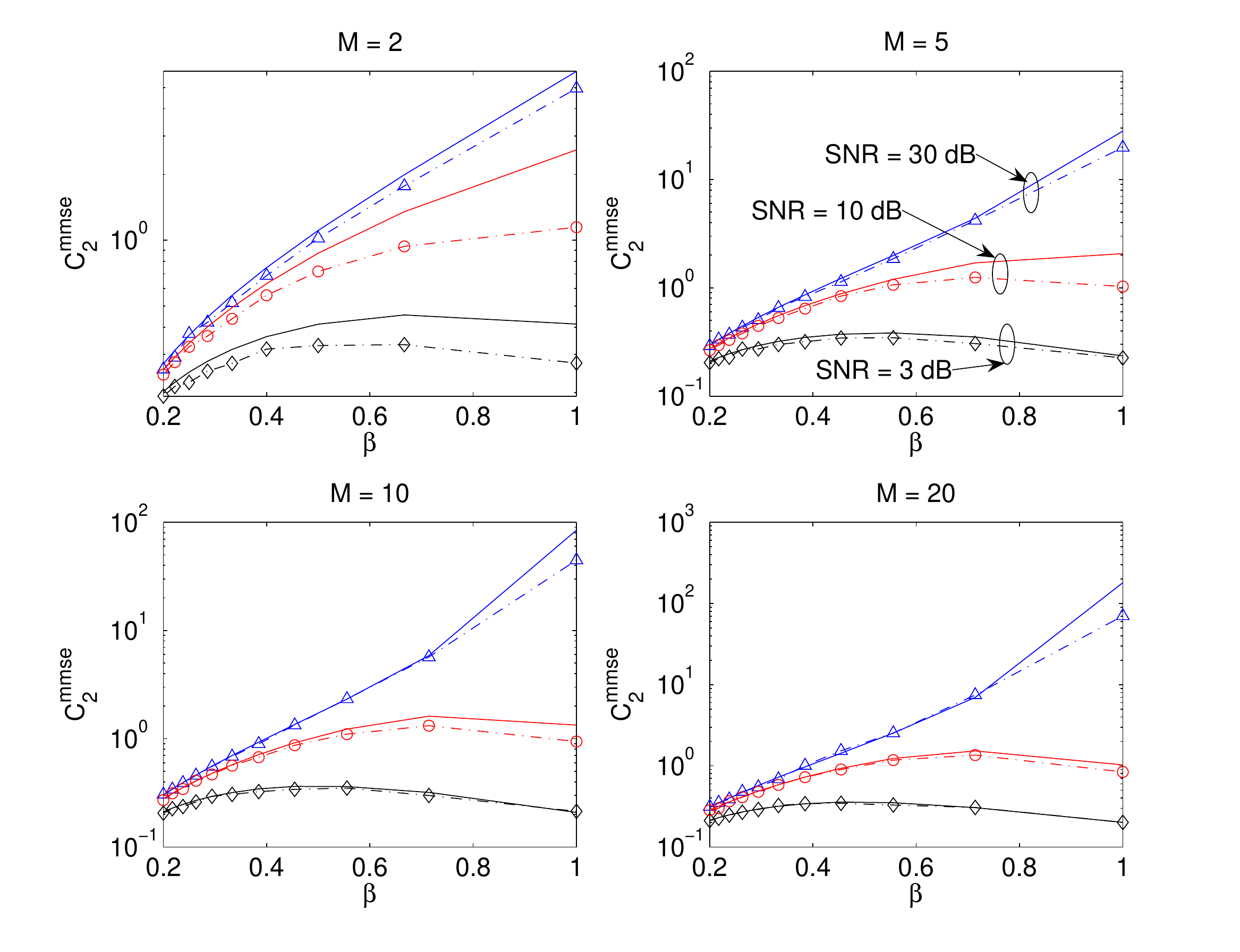}
\caption{Variance of the MMSE mutual information (${\cal
C}_2^{\mmse}$) as a function of $\beta$, for $M = 2,5,10$ and $20$.
The solid lines are analytical results, and the corresponding
dash-dot lines are empirical results obtained from Monte Carlo
simulation. Diamonds denote 3 dB, circles 10 dB and triangles 30
dB.} \label{fig:MMSE_variance}
\end{center}
\end{figure}

\begin{figure}[htbp]
\begin{center}
\hspace*{-0.65in}
\includegraphics[width=21cm]{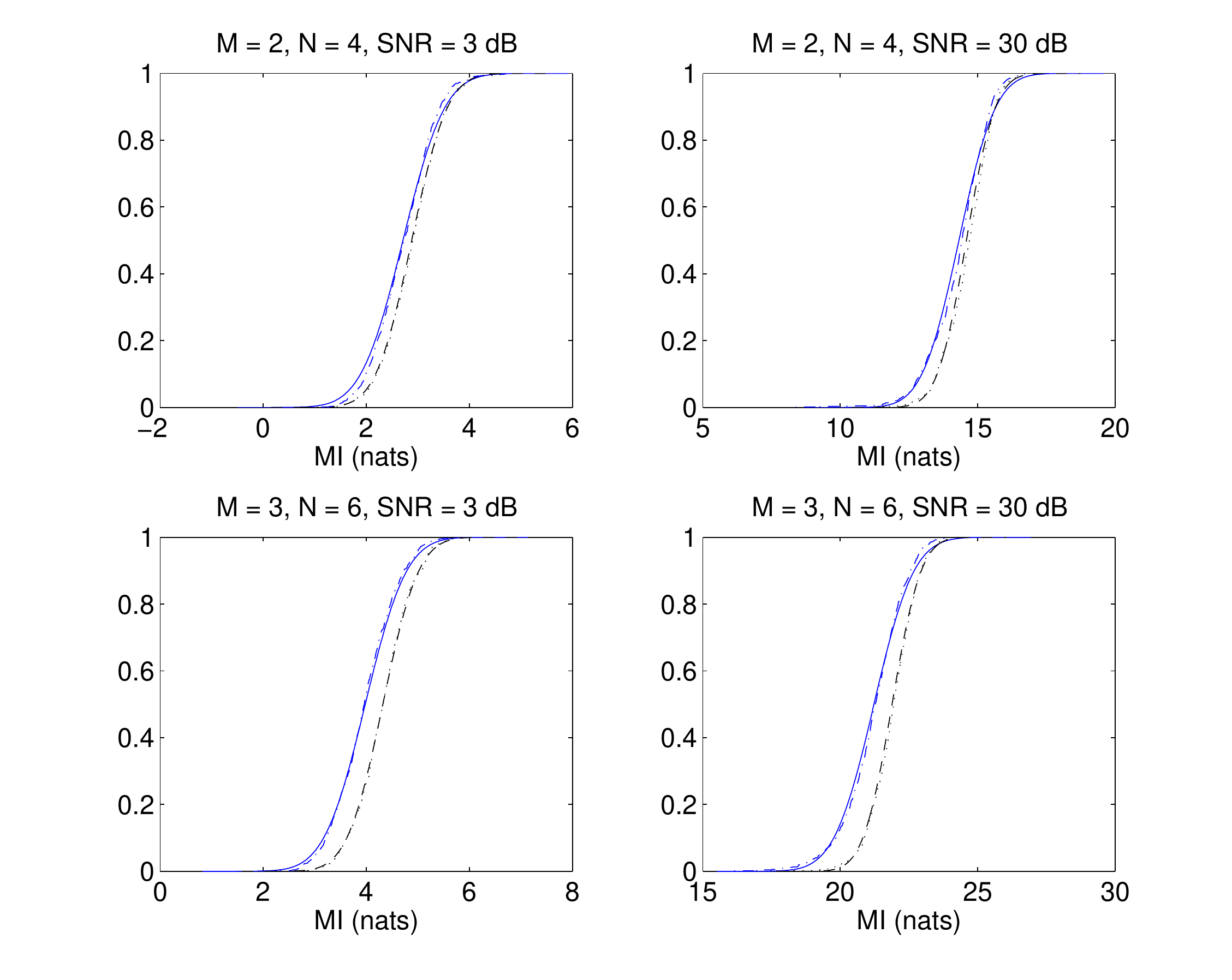}
\caption{CDF of the mutual information (MI) for the MMSE and optimal
receivers, for $M = 2,3$, $\beta = 0.5$ and $\rho = 3, 30$ dB. The
solid blue line is the analytical result for the MMSE, the dot-dash
blue is MMSE empirical, the dashed black is Optimal receiver
analytical and the dotted black is the optimal receiver empirical.}
\label{fig:CDF_MI_2}
\end{center}
\end{figure}

\begin{figure}[htbp]
\begin{center}
\hspace*{-0.65in}
\includegraphics[width=21cm]{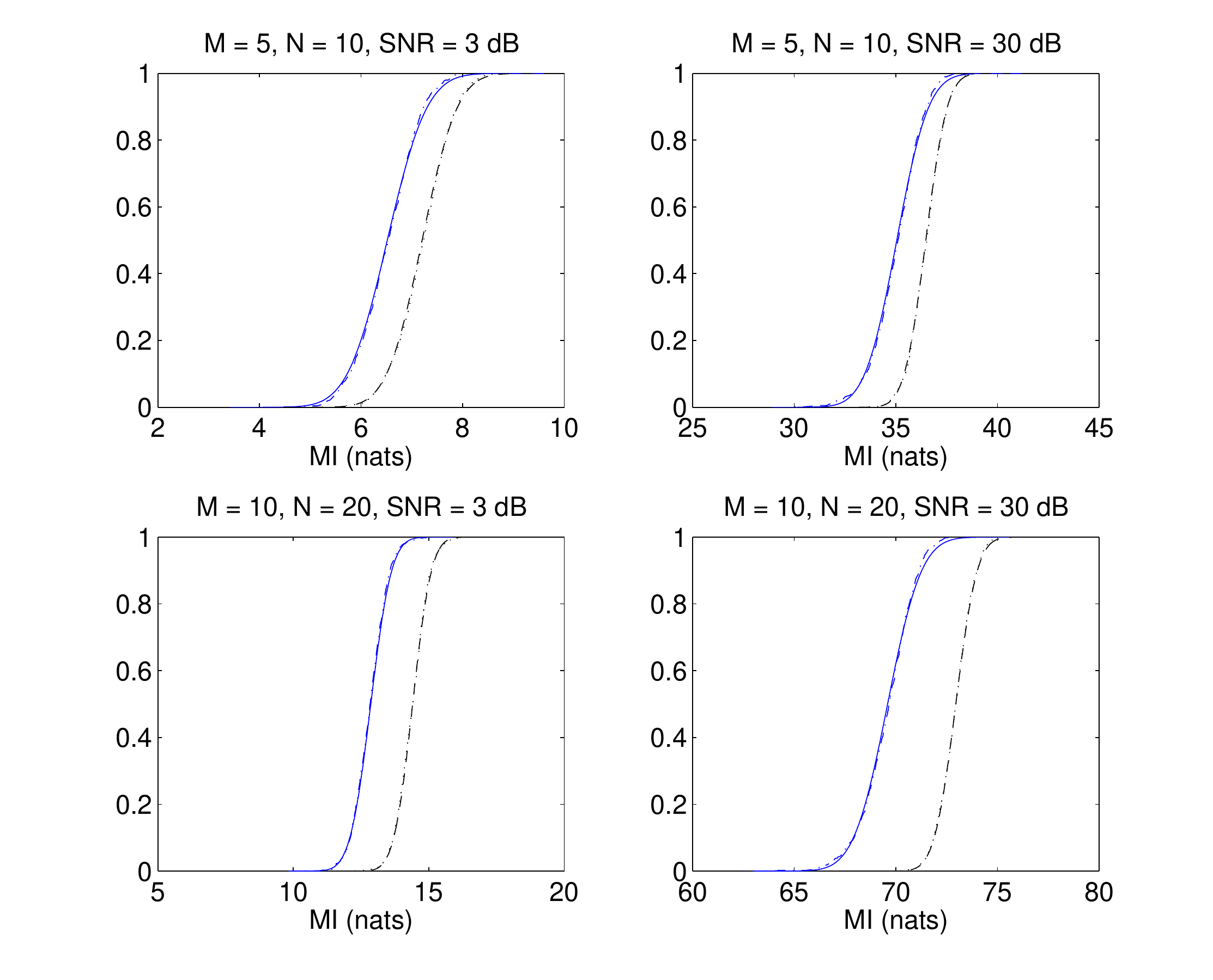}
\caption{CDF of the mutual information (MI) for the MMSE and optimal
receivers, for $M = 5,10$, $\beta = 0.5$ and $\rho = 3, 30$ dB. The
solid blue line is the analytical result for the MMSE, the dot-dash
blue is MMSE empirical, the dashed black is Optimal receiver
analytical and the dotted black is the optimal receiver empirical.}
\label{fig:CDF_MI}
\end{center}
\end{figure}

\begin{figure}[htbp]
\begin{center}
\hspace*{-0.65in}
\includegraphics[width=21cm]{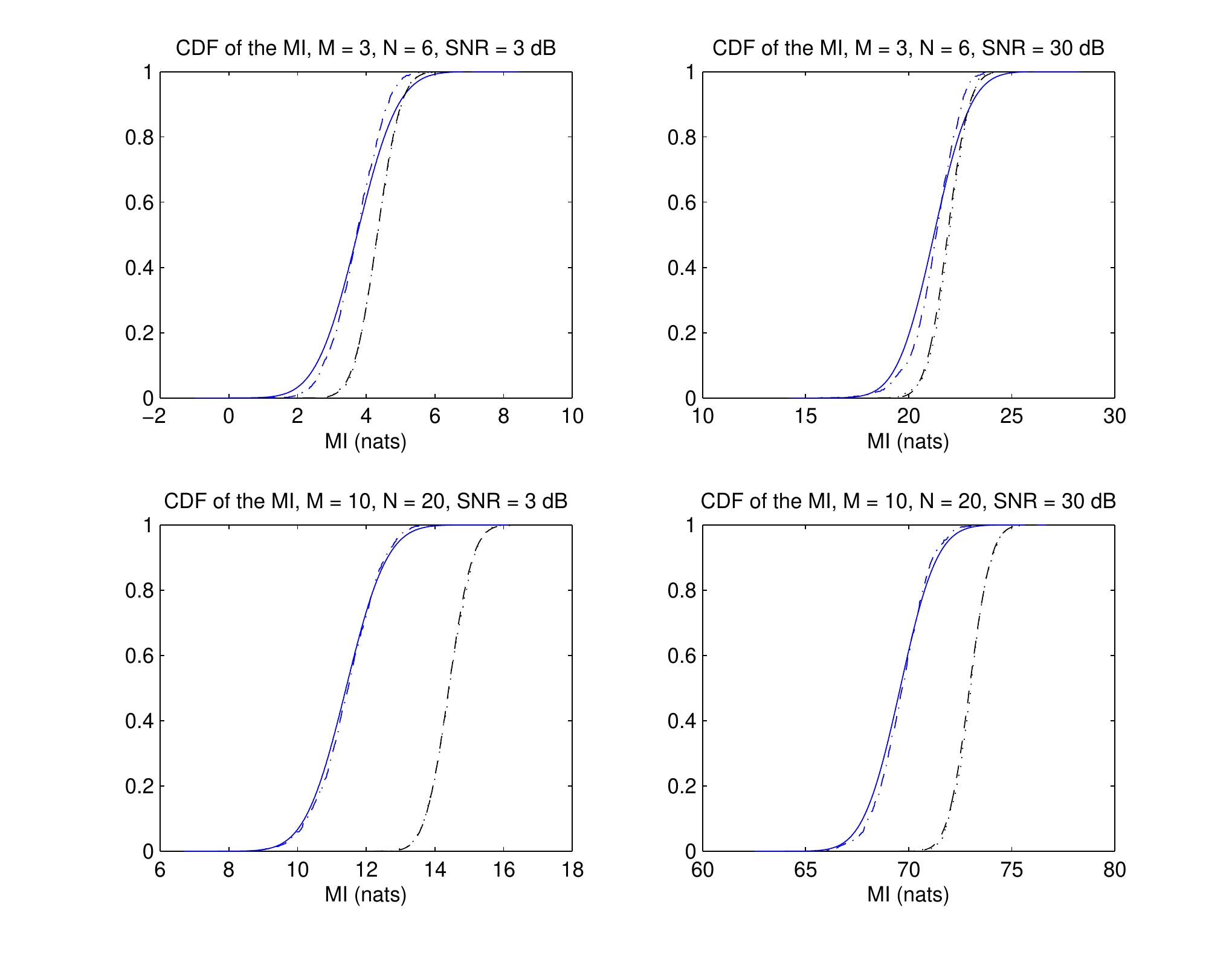}
\caption{CDF of the mutual information (MI) for the ZF and optimal
receivers, for $M = 3,10$, $\beta = 0.5$ and $\rho = 3, 30$ dB. The
solid blue line is the analytical result for the ZF receiver, the
dot-dash blue is ZF empirical, the dashed black is Optimal receiver
analytical and the dotted black is the optimal receiver empirical.}
\label{fig:CDF_MI_ZF}
\end{center}
\end{figure}

We conclude this section by presenting some numerical results.
Fig.~\ref{fig:MMSE_mean} compares the analytical mean of the MMSE mutual information per antenna ${\cal C}_1^\mmse/M$ with the
corresponding empirical mean obtained from Monte Carlo simulation.
The corresponding comparison between the analytical variance ${\cal
C}_2^\mmse$ and the empirical variance is presented in
Fig.~\ref{fig:MMSE_variance}.
Using the results for the mean and the variance, we plot the CDF of
the (Gaussian) mutual information for the MMSE and optimal receiver
in Figs.~\ref{fig:CDF_MI_2},\ref{fig:CDF_MI}. Both analytical and
empirical results are plotted, for a wide range of $M,N$ and SNRs.
For brevity, we have only reported plots of the CDF for
the mutual information for the ZF case, see
Fig.~\ref{fig:CDF_MI_ZF}. The plots are for $M = 3, 10$, $\beta =
0.5$ and $\rho = 3, 30$ dB. The results are similar in flavor to the
MMSE case.

We notice that the analytical and empirical results match closely,
for even moderate number of antennas and not too large SNRs,
in line with the comments made earlier regarding the validity of the
analysis.  It is also worthwhile noticing that the accuracy of the
Gaussian approximation for linear receivers appears to be slightly
inferior to that of the  Gaussian approximation for the optimal receiver case,
especially for very small $N$ and large SNR.

\section{Conclusions} \label{conclusions}

Novel wireless communication systems are targeting very large spectral
efficiencies and will operate at high SNR thanks to hot-spots and pico-cell arrangements.
For example, a system with bandwidth of 20 MHz and operating at 100 Mb/s requires a
spectral efficiency of 5 bit/s/Hz, corresponding to coding rate $R = 5$ bpcu in notation adopted here,
if one neglects non-ideal effects such as pilot symbols, guard band and guard
intervals, cyclic prefix redundancy for OFDM, etc.
For such systems, the use of low-complexity
linear receivers in a {\em separated detection and decoding}
architecture as those examined in this paper may be
mandatory because of complexity and power consumption.

In this paper we investigated the asymptotic performance of such
separated linear detection and decoding architectures in two
relevant asymptotic regimes. In the regime of fixed number of
antennas and increasing SNR and coding rate, we showed that linear
detection may be very suboptimal. Furthermore, due to the strong
correlation between the SINRs of the parallel channels induced by
the linear receiver, coding across the antennas does not help in
terms of the achievable Diversity-Multiplexing Tradeoff. We also
illuminated the very peculiar behavior of the linear MMSE receiver
with coding across the antennas, that exhibits a diversity order
(slope of the outage probability curve) that changes depending on
the rate. Then, we analyzed the asymptotic behavior of the linear
MMSE and the ZF receivers with coding across the antennas
in the regime of fixed SNR and large (but finite) number of
antennas. We showed that the corresponding mutual information
has statistical fluctuations that converge in distribution to a Gaussian random variable, and we
computed its mean and variance in closed form.
This yields a simple Gaussian approximation of the outage probability in
this asymptotic regime,  within the limitations that have been thoroughly discussed.

Based on the analysis carried out in this work, we may summarize
some considerations on system design. In order to achieve a
required target spectral efficiency at given block-error rate and SNR operating point, an
attractive design option may consists of increasing the number of
antennas (especially at the receiver) and using a low-complexity
linear receiver. However, pure spatial multiplexing (independent
coded streams directly fed into the transmit antennas)
and/or linear ZF receivers should be avoided. In contrast, coding across
antennas and a linear MMSE receiver can achieve a very good
tradeoff between performance and complexity in a wide range of
system operating points.

\clearpage
\newpage

\appendix

\centerline{APPENDIX}

\section{Proof of $P(\Ac) = O(1)$} \label{appendix-A}

\begin{figure}[htbp]
\begin{center}
\includegraphics[width=12cm]{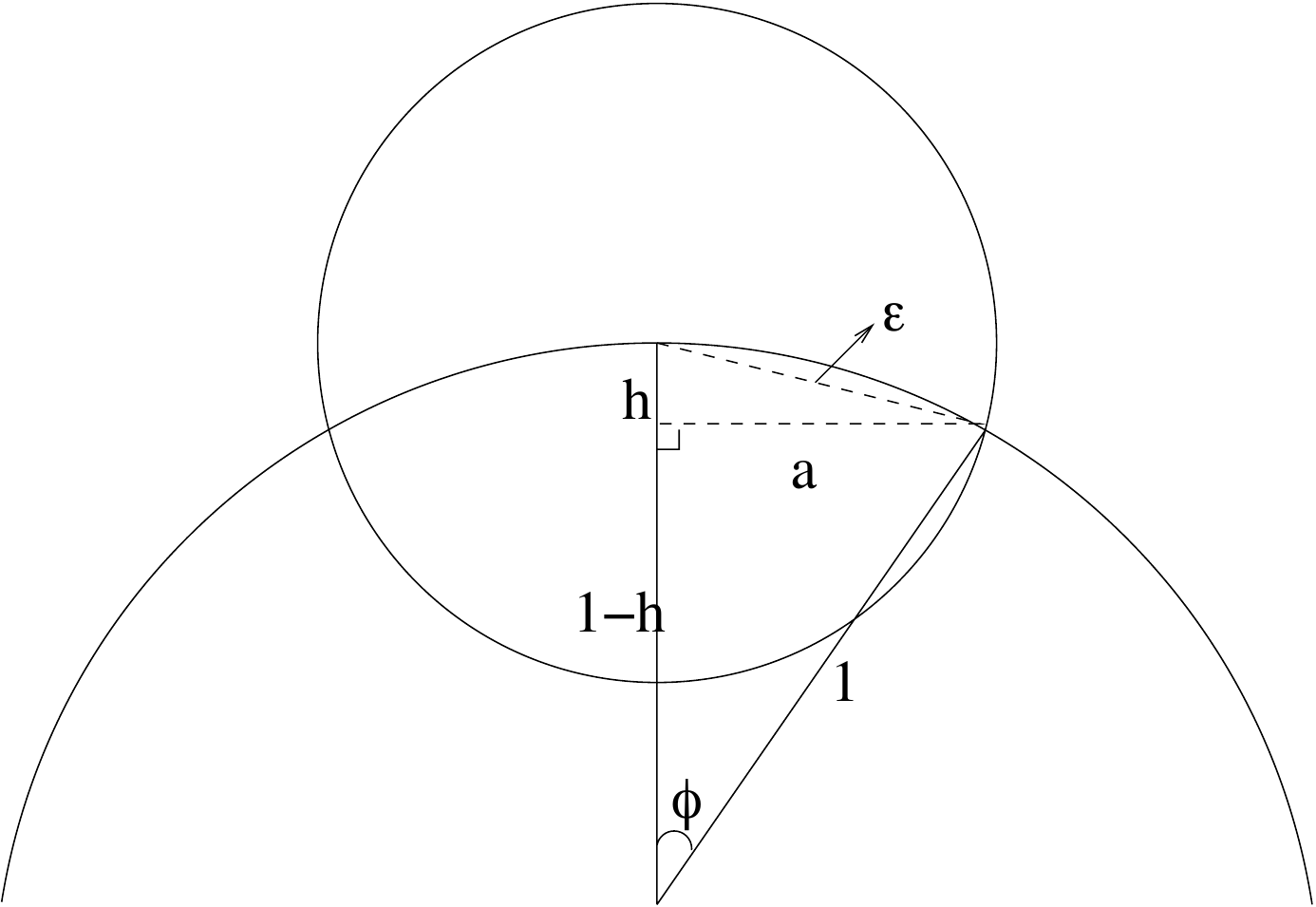}
\caption{The unit hemisphere and a spherical cap.}
\label{fig:sph_cap}
\end{center}
\end{figure}

Our aim is to provide a lower-bound to the quantity
\[ P ( \mathcal{A} ) = P \left ( \frac{1}{M}
\sum_{k=1}^{M} \frac{1}{|u_{1 k}|^2} \leq c \right ).\] It is well
known that $\Um$ and $\Lambdam$ are independent random matrices
and that $\Um$ is Haar distributed (i.e., is an isotropically
random unitary matrix distributed uniformly over the Stiefel
manifold).
Therefore, the vector $\uv_1 \triangleq
(u_{1,1},u_{1,2},\hdots,u_{1,M})$ corresponding to the first row
of $\Um$ is uniformly distributed on the unit $M$-dimensional
hypersphere (or $M$-sphere) $\mathcal{O}_M(\zerov,1)$,\footnote{We
will use the notation $\mathcal{O}_M(\cv,\delta)$ to denote an
$M$-sphere centred at $\cv$ with radius $\delta$.} and satisfies
$|\uv_1|^2 = \sum_{i=1}^{M} |u_{1 i}|^2 = 1$. The point $\pv =
\left(
\frac{1}{\sqrt{M}},\frac{1}{\sqrt{M}},\hdots,\frac{1}{\sqrt{M}}
\right)$ is a point on the unit $M$-sphere. Consider the spherical
cap $\mathcal{C}$ of the unit sphere that is cut out by the sphere
$\mathcal{O}_M(\pv,\epsilon)$, where $\epsilon$ is a small
positive number (See Fig.~\ref{fig:sph_cap}).

The coordinates of any point $\uv$ in this spherical cap
$\mathcal{C}$ is lower bounded by
\[ u_{j} \geq \frac{1}{\sqrt{M}} - \epsilon, \ \forall \ j = 1,\hdots,M. \]
Therefore,
\[ \frac{1}{M} \sum_{j=1}^{M} \frac{1}{|u_{j}|^2} \leq \frac{1}{\left( \frac{1}{\sqrt{M}} - \epsilon \right)^2}.\]
Defining the constant $c = \left( \frac{1}{\sqrt{M}} - \epsilon
\right)^{-2}$, we have that
\begin{eqnarray*}
P ( \mathcal{A} ) &\geq& P \left ( \uv_1 \in \mathcal{C}  \right ) \\
&=& \frac{\text{Surface area of }\mathcal{C}}{\text{Surface area
of }\mathcal{O}_M(\zerov,1)},
\end{eqnarray*}
where the latter equality holds since $\uv_1$ is isotropic. In
order to compute the above surface areas, consider
Fig.~\ref{fig:sph_cap}. The surface area of $\mathcal{C}$, denoted
by $\Omega(\phi)$ is given by \cite{Sha}
\[ \Omega(\phi) = \frac{(M-1)\pi^{(M-1)/2}}{\Gamma \left( \frac{M+1}{2} \right)} \int_0^\phi (\sin \theta)^{M-2} d\theta, \]
and the surface area of an unit $M$-sphere is \cite{Sha} $S_M(1) =
M \pi^{M/2}/\Gamma(M/2 + 1)$. All that remains is to compute the
angle $\phi$, which is accomplished by solving the following
equations obtained from the two right-angled triangles in
Fig.~\ref{fig:sph_cap}:
\begin{eqnarray*}
h^2 + a^2 &=& \epsilon^2\\
a^2 + (1-h)^2 &=& 1.
\end{eqnarray*}
Solving for $a,h$, we obtain
\[ h = \frac{\epsilon^2}{2}, \ a = \epsilon \sqrt{1 - \frac{\epsilon^2}{4}}.\]
Therefore,
\[ \phi = \tan^{-1} \left( \frac{\epsilon \sqrt{1 - \frac{\epsilon^2}{4}}}{1-\frac{\epsilon^2}{2}} \right),\]
and
\[ P ( \mathcal{A} ) \geq \frac{\Omega(\phi)}{S_M(1)} > 0,\]
as desired.


\section{Novikov's Theorem} \label{app:novikov_thm}

We introduce here a useful trick which makes the connection
between Gaussian integration and differentiation over Gaussian
random variables.

\begin{thm}\label{thm:Novikov}\cite{Novikov1965_NovikovTheorem}
(Novikov) Let $\Hm$ be an $N \times M$ matrix with i.i.d. elements
drawn from $\Cc\Nc(0,1/N)$. Let $f(\Hm,\Hm^\herm)$ be a scalar
function of the matrix elements of $\Hm$ and their complex
conjugates, which does not grow faster than the inverse of the
probability density of $\Hm$, i.e., such that
\begin{equation}\label{eq:novikov_thm_condition}
\lim_{\left|h_{ij}\right|\rightarrow \infty}p(\Hm)f(\Hm,\Hm^H)=0
\end{equation}
for all matrix elements $h_{ij}$. Then, for any set of indices
$i,j$, the following relation holds
\begin{eqnarray}\label{eq:novikov_thm}
\EE\left[h_{ij}f(\Hm,\Hm^\herm)\right] &=&
\frac{1}{N}\EE\left[\frac{\partial f(\Hm,\Hm^\herm)}{\partial
h^*_{ij}}\right],
\end{eqnarray}
where $h_{ij}$ and $h^*_{ij}$ are to be treated as individual
variables in the differentiation.\footnote{This means that
$\frac{\partial h_{i,j}^*}{\partial h_{i,j}} = 0$ when computing the
partial derivative in (\ref{eq:novikov_thm}).}
\end{thm}

{\bf Sketch of the Proof.} Even though the general proof is
involved, we present here a simple proof for a single real Gaussian
variable ${\cal N}(0,1)$ by integrating by parts:
\begin{eqnarray}\label{eq:novikov_thm_example}
\EE\left[z f(z) \right] &=& \int_{-\infty}^\infty
\frac{dz}{\sqrt{2\pi}} z f(z)e^{-\frac{z^2}{2}} \\
\nonumber %
&=& - \left.\frac{e^{-\frac{z^2}{2}} f(z)}{\sqrt{2\pi }}
\right|_{-\infty}^\infty + \int_{-\infty}^\infty
\frac{dz}{\sqrt{2\pi}} f'(z)e^{-\frac{z^2}{2}} \\
\nonumber &=& \EE\left[f'(z) \right]
\end{eqnarray}
\hfill $\square$

{\bf Example.}
A useful and illustrative example of the application of Novikov Theorem
is to evaluate the fourth moment of a i.i.d. complex Gaussian vector
with elements drawn from $\Cc\Nc(0,1/N)$. We have
\begin{eqnarray}
    \label{eq:novikov_thm_2}
    \EE\left[ h_{i}^* h_{j} h_{k}^* h_{l} \right] &=&
    \frac{1}{N} \EE\left[ \frac{\partial}{\partial h_{i}} \left( h_{j} h_{k}^*
    h_{l} \right)
    \right] =
    \frac{\delta_{ij}}{N} \EE\left[ h_{k}^* h_{l} \right] +
    \frac{\delta_{il}}{N} \EE\left[ h_{k}^* h_{j} \right] \\
    \nonumber
    &=&
    \frac{\delta_{ij} \delta_{kl}}{N^2} +     \frac{\delta_{il} \delta_{kj}}{N^2}
\end{eqnarray}

\section{Fluctuations of eigenvalues} \label{app:eig_flucts}

Let
\begin{equation}\label{eq:A_mat_def}
\Bm  = \left ( \Id + \alpha\Hm\Hm^\herm \right )^{-1}
\end{equation}
where $\Hm \in \CC^{N \times M}$ has i.i.d. elements
$\Cc\Nc(0,1/N)$, and let $\beta = M/N$. The normalized trace of
$\Bm$ is given by
\[ \frac{1}{N} {\rm Tr} \left (\Bm \right ) = \frac{1}{N} \sum_{k=1}^N \frac{1}{1 + \alpha
\lambda_k(\Hm\Hm^\herm)} =  \eta^{(N)}(\alpha) \] %
where $\eta^{(N)}(\alpha)$ is the $\eta$-transform
\cite{Tulino2004_RMTInfoTheoryReview} of the empirical eigenvalue
distribution of $\Hm\Hm^\herm$. For large $N$, the variance of ${\rm
Tr} \left (\Bm \right ) = N \eta^{(N)}(\alpha)$ is of order unity
and can be calculated in closed form
\cite{Beenakker1997_MesoscopicReview, Politzer1989_RMTCumulants},
\begin{eqnarray}
\label{eq:cum_moment2_eigs1} %
\EE_c\left[{\rm Tr}(\Bm) ; {\rm Tr}( \Bm)\right] &=& {\cal P}
\int_{\lambda_{min}}^{\lambda_{\max}} d\lambda\,
 \int_{\lambda_{min}}^{\lambda_{\max}} d\mu\,
\Kc_2(\lambda,\mu) \frac{1}{(1+\alpha\lambda)(1+\alpha\mu)},
\end{eqnarray}
where ${\cal P}$ denotes the principal part of the integral,
$\lambda_{min,max}$ denote the extremal values of the support of the
Marcenko-Pastur law \cite{Tulino2004_RMTInfoTheoryReview}, given by
\begin{eqnarray}\label{eq:lambda_minmax}
\lambda_{min,max} = \left(1\pm \sqrt{\beta}\right)^2,
\end{eqnarray}
and $\Kc_2(\lambda,\mu)$ is an integral kernel representing the
deviation of the joint eigenvalue distribution of the eigenvalues
$\lambda$, $\mu$ from the product of their marginal distributions
and is given asymptotically by
\begin{eqnarray}
\label{eq:cum_moment2_K2} %
\Kc_2(\lambda,\mu) &=& \frac{1}{2\pi^2} \frac{1}{\sqrt{
\left(\lambda - \lambda_{min}\right)\left(\lambda_{max} -
\lambda\right)}} \frac{\partial}{\partial\mu}
 \left[\frac{\sqrt{\left(\mu -
\lambda_{min}\right)\left(\lambda_{max} - \mu\right)}}{\lambda -
\mu}\right].
\end{eqnarray}
It can be verified that the above function is symmetric in $\lambda$
and $\mu$. After integration by parts, we get
\begin{eqnarray}
\label{eq:cum_moment2_eigs} %
\EE_c\left[{\rm Tr}(\Bm) ; {\rm Tr}( \Bm)\right] &=&
\frac{1}{2\pi^2} {\cal P} \int_{\lambda_{min}}^{\lambda_{\max}}
d\lambda\,
 \int_{\lambda_{min}}^{\lambda_{\max}} d\mu\,
\left[\frac{\left(\mu - \lambda_{min}\right)\left(\lambda_{max} -
\mu\right)}{\left(\lambda -
\lambda_{min}\right)\left(\lambda_{max} -
\lambda\right)}\right]^{1/2} \frac{\alpha}{\left(\mu -
\lambda\right)\left(1+\alpha\lambda\right)\left(1+\alpha\mu\right)^2}
 \nonumber \\
&=&
\frac{\alpha^2\beta}{\left(1+2\alpha(1+\beta)+\alpha^2(1-\beta)^2\right)^2}
= O(1).
\end{eqnarray}
The result (\ref{eq:cum_moment2_eigs}) is used in the calculation
of the correlations of $\gamma_1$, $\gamma_2$ in
(\ref{gamma_xcorr1_6}).

Furthermore, in \cite{Beenakker1997_MesoscopicReview,
Politzer1989_RMTCumulants} two important and more general results
are shown. In particular, for any functions $f_1(x)$, $f_2(x)$ the
following result is true in the limit of large $N$
\begin{eqnarray}
\label{eq:beenakker_gen_result2} %
\EE_c\left[{\rm Tr}(f_1(\Bm)) ; {\rm Tr}( f_2(\Bm))\right] &=& {\cal P}
\int_{\lambda_{min}}^{\lambda_{\max}} d\lambda\,
 \int_{\lambda_{min}}^{\lambda_{\max}} d\mu\,
\Kc_2(\lambda,\mu) f_1(\lambda) f_2(\mu) = O(1)
\end{eqnarray}
as long as these functions are bounded and smooth enough within the
support of the asymptotic eigenvalue spectrum (for example
$f(x)=[x]_+$ is not smooth, while $f(x)=\alpha/(1+\alpha x)^2$ is
smooth).

Also, in \cite{Beenakker1997_MesoscopicReview,
Politzer1989_RMTCumulants} it is shown that all higher order
cumulants of such smooth functions vanish in the large $N$ limit,
i.e., for $n>2$
\begin{equation}\label{eq:beenakker_gen_result3}
  \EE_c\left[{\rm Tr} \left( f_1(\Hm\Hm^\herm) \right);\ldots; {\rm Tr} \left( f_n(\Hm\Hm^\herm) \right)\right] = {\cal Q}_{n}(N) = o(1)
\end{equation}
where we have denoted the (arbitrary for our purposes) scaling of
the above cumulant moment with $N$ as ${\cal Q}_{n}(N)$ for future
reference. We will use this result in Appendix~\ref{app:asympt_cum}
to prove that all cumulant moments of $I_N$ of order higher than two
vanish for large $N$.

\section{Higher order cumulants are vanishing} \label{app:asympt_cum}

After the calculation of the mean (\ref{eq:C_1_mmse}) and the
variance (\ref{eq:Var_MI_MMSE}) of the mutual information in Section
\ref{sec:MMSE_MI_distrib} we now need to show that the higher order
cumulants of the mutual information vanish in the limit of large
$N$. This will conclude the proof of the asymptotic Gaussianity of
the mutual information, as discussed in the beginning of Section
\ref{sec:MMSE_MI_distrib}.

We need to show that the cumulant moments defined in (\ref{eq:cum_product_def}) as
\begin{eqnarray*}\label{eq:app_cum_product_def}
{\cal C}_n = \sum_{k_1,\ldots
k_n=1}^M\EE_c[\log(1+\gamma_{k_1});\ldots;\log(1+\gamma_{k_n})].
\end{eqnarray*}
vanish for $n>2$ when $N\rightarrow \infty$.
Despite the fact that ${\cal C}_n$ is a sum of $O(N^n)$ terms,
we shall show that it is in fact of order $o(1)$.

\subsection{MMSE receiver higher order cumulants}

We discuss in some detail the case of the MMSE receiver.
At the end of this appendix, a short argument is given in order to reach the same
conclusions for the ZF case.

While a formal proof would be lengthy and tedious and would not add
much value to the paper,  we shall provide a  sketch the basic steps
of the proof leaving out several technicalities. We start by
recalling that each $\gamma_{k_i}$ in the above sum is defined as
$\gamma_{k_i}=\alpha\hv_{k_i}^\herm \Bm_{k_i} \hv_{k_i}$, where
$\Bm_{k_i} = \left (\Id
+\alpha\Hm\Hm^\herm-\alpha\hv_{k_i}\hv_{k_i}^\herm \right )^{-1}$.
Using the matrix inversion lemma, we see that the denominator in the
expression of $\Bm_{k_i}$ includes all columns of $\Hm$ other than
$\hv_{k_i}$. For every $n$-tuple $\{k_1,k_2,\ldots,k_n\}$ we define
the matrix $\Bm_{\{k_i\}}$, such that
\[ \Bm_{\{k_i\}} = \left ( \Id+\alpha\Hm_{\{k_i\}}\Hm_{\{k_i\}}^\herm \right )^{-1}, \]
where $\Hm_{\{k_i\}}$ is obtained by removing columns
$\hv_{k_1},\hv_{k_2},\ldots,\hv_{k_n}$ from $\Hm$. If for some
$i\neq j$ we have $k_i=k_j$, then we only remove column $k_i$ once.
For any finite $n$, the following approximation holds
\begin{equation}\label{eq:app_approx_B_ki_B_kall}
  \Bm_{k_i} = \Bm_{\{k_i\}} + O\left(\frac{1}{N}\right),
\end{equation}
in the sense that the elements of the matrix $N(\Bm_{k_i} -
\Bm_{\{k_i\}})$ are almost surely finite in the limit of large $N$.
Roughly speaking, since the elements of $\Hm$ are zero-mean Gaussian with variance $1/N$, the difference
$\Bm_{k_i}^{-1} - \Bm_{\{k_i\}}^{-1} = \alpha\sum_{k_j\neq k_i}
\hv_{k_j}\hv_{k_j}^\herm$ adds a term of order $O(1/N)$ to each
element of $\Bm$, which can be neglected for large $N$. For example,
of the above approximation can be proved in an iterative fashion,
by showing that $\Bm_{k_1} - \Bm_{k_1,k_2}$ is $O(1/N)$,
then adding to writing $\Bm_{k_1}=\Bm_{k_1}-\Bm_{k_1,k_2}+\Bm_{k_1,k_2} - \Bm_{k_1,k_2,k_3}$
and then showing that $\Bm_{k_1,k_2}-\Bm_{k_1,k_2,k_3}$ is also
$O(1/N)$ etc. Each such difference can be shown to be almost
surely $O(1/N)$ by applying the matrix inversion lemma and
observing that the elements of $\hv_{k_i}$ are ${\cal CN}(0,1/N)$.

As a result, to leading order in $1/N$, we have
\begin{eqnarray*}\label{eq:app_cum_product_def2}
{\cal C}_n \approx \sum_{\{k_i\}} \EE_c[
\log(1+\alpha\hv_{k_1}^\herm\Bm_{\{k_i\}}\hv_{k_1}); \ldots;
\log(1+\alpha\hv_{k_n}^\herm\Bm_{\{k_i\}}\hv_{k_n})]
\end{eqnarray*}
where $\sum_{\{k_i\}}$ is a sum over all possible $1\leq k_1,\ldots,k_n\leq M$.

Since now the random vectors $\hv_{k_1},\ldots,\hv_{k_n}$ do not
appear in $\Bm_{\{k_i\}}$, we can explicitly average over them in
the above expression. At this point it is convenient to expand the
logarithms in Taylor series, such that each term in the sum above becomes
\begin{eqnarray}\label{eq:app_cum_product_def3}
\sum_{l_1,\ldots,l_n=1}^\infty
\frac{(-\alpha)^{l_1+\ldots+l_n}}{l_1 l_2\ldots l_n}
\sum_{\{k_i\}}
\EE_c\left [ \left (\hv_{k_1}^\herm\Bm_{\{k_i\}}\hv_{k_1} \right )^{l_1};\ldots;\left ( \hv_{k_n}^\herm\Bm_{\{k_i\}}\hv_{k_n} \right )^{l_n} \right ].
\end{eqnarray}
We may now apply Novikov Theorem to average over the $\hv_{k_i}$'s.
This is in general a formidable exercise in combinatorics. Instead,
we only need to find how the leading terms scale with $N$.
Specifically, in the following we will fix the $n$-tuple
$l_1,l_2\,\ldots,l_n$ and show that the corresponding term of the
type $\sum_{\{k_i\}}\EE_c[\cdot ]$ is $o(1)$ as $N\rightarrow
\infty$.

We first notice that since there are $L_{\{l_i\}} = l_1+\ldots
+l_n$ pairs of $\hv^\herm$'s and $\hv$'s,
we will have an overall factor of $N^{-L_{\{l_i\}}}$ after averaging
over all $\hv$'s. Also, we can decompose the sum over
$\{k_i\}$ into subsets or ``shells'', where each shell has the same number of
distinct indices $k_i$'s. For example, there are ${M \choose n} = O(N^n)$ terms containing all
distinct indices, and $nM!/((M-n+1)!(n-1)!) = O(N^{n-1})$ terms having
$n-1$ distinct indices and one repeated index.
In general, the shell with $q$ distinct indices contains $O(N^q)$ terms.

If the $k_i$'s are all distinct, then the resulting cumulant moment will be of order $n$.
Possible terms that may appear include, for example,
\begin{eqnarray}\label{eq:app_cum_product_def4}
\EE_c[{\rm Tr} \left[\Bm_{\{k_i\}}\right]^{l_1}; \ldots; {\rm Tr}
\left[\Bm_{\{k_i\}}\right]^{l_n}]\\ \nonumber
\EE_c[{\rm Tr} \left[\Bm_{\{k_i\}}\right]^{l_1-2} {\rm Tr}
\left[\Bm_{\{k_i\}}^2\right]; \ldots; {\rm Tr}
\left[\Bm_{\{k_i\}}\right]^{l_n}]
\end{eqnarray}
where a term as in the second line can only appear for
$l_1\geq 2$.
Therefore, after averaging over the
$\hv_{k_i}$'s we are left with order-$n$ moments,
having as arguments products of the random variables ${\rm Tr}\Bm_{\{k_i\}}^{m_j}$, for
$1\geq m_j \geq l_j $, with $j = 1,\ldots,n$.  Let $x_j$ be the total number of traces appearing in
the $j$-th argument of a particular term, such that $x_j \leq
l_j$.
For example, in the first line of (\ref{eq:app_cum_product_def4}), we have $x_1 = 1$, while
in the second line we have $x_1 = 2$.
Since generally $x_i\geq 1$, such cumulant moments can be
reduced to sums of products of irreducible moments with respect to
these random variables. To estimate the leading scaling in $N$ of
these reducible moments, we recall that the first moment of the
trace is $\EE_c [ {\rm Tr} \Bm_{\{k_i\}}^m] = O(N)$, the second
cumulant moment is $O(1)$ (\ref{eq:beenakker_gen_result2}), while
all higher cumulant moments are $o(1)$
(\ref{eq:beenakker_gen_result3}).
Let the leading term in the expansion of the reducible moments into irreducible ones have
$d_1$ cumulants of order one, $d_2$ cumulants of order 2,
$d_3$ cumulants of order 3 etc. The only constraint we need to impose is
that %
\[d_1 \leq d_{1max} = \sum_{i=1}^n \left[x_i - 1\right]_+ \leq L_{\{l_i\}} - n, \]
which is valid due to the shift-invariance of irreducible cumulant
moments of order higher than one. In the case that $d_1=d_{1max}$,
we need to have $d_n = 1$ and $d_k=0$ for all $k\neq 1,n$.
Collecting all powers of $N$ this term will be of order
$N^{n-L_{\{l_i\}}+d_{1max}}{\cal Q}_{n}(N)\leq {\cal Q}_{n}(N) =
o(1)$, where we recall that the factor $N^n$ is due to the
$O(N^n)$ possible combinations of the distinct $k_i$'s that appear
in the corresponding sum in (\ref{eq:app_cum_product_def3}), while
the factor $N^{-L_{\{l_i\}}}$ comes from the averages over the
$\hv_{k_i}$. Otherwise, when $d_1<d_{1max}$, the leading term will
have, if that is at all possible, $d_2=(\sum_i x_i - d_1)/2$ and
$d_k=0$ for $k>2$. In this case, the scaling of this term with $N$
is $N^{n-L_{\{l_i\}}+d_{1}}\leq N^{-1} = o(1)$.

The above argument can be extended to the case when there are
$q<n$ distinct $k_i$'s. The difference is that now additional
terms may appear, since Novikov's formula may give derivatives
across different arguments of the cumulant moment, thus reducing
the order of the cumulant, e.g. if $k_1=k_2$ we will get the term
\begin{eqnarray}\label{eq:app_cum_product_def5}
\EE_c[{\rm Tr} \left[\Bm_{\{k_i\}}\right]^{l_1-1} {\rm Tr}
\left[\Bm_{\{k_i\}}\right]^{l_2-1} {\rm Tr}
\left[\Bm_{\{k_i\}}^2\right]; {\rm Tr}
\left[\Bm_{\{k_i\}}\right]^{l_3} \ldots; {\rm Tr}
\left[\Bm_{\{k_i\}}\right]^{l_n}].
\end{eqnarray}
In general, for the shell with $q < n$ distinct indices the resulting terms
will include cumulant moments with orders $m$, such that $q\leq m\leq n$. When $m=n$, we
can directly apply the argument used when $q=n$,
only replacing the number of possible combinations of the distinct combinations
from $N^n$ to $N^q$.

In the case that $m < n$,  when one expresses the reducible cumulant
moments in terms of sums over products of irreducible ones, the
maximum number of order one cumulants that may appear is now
bounded by
\[d_1 \leq d_{1max}' = \sum_{i=1}^m \left[x_i - 1\right]_+ < L_{\{l_i\}} - q. \]
The crucial difference is that $d_{1max}'< L_{\{l_i\}} - q$, which
is due to the fact that, as seen in
(\ref{eq:app_cum_product_def5}), in order to reduce the order of
the cumulant moment to $m<n$, one needs to produce traces that
span different arguments of the original cumulant moments. In this
case, the leading term will be of order $N^{q-L_{\{l_i\}}+d_{1}}\leq N^{q-L_{\{l_i\}}+d_{1max}'} = o(1)$.

Following the above argument, we can show that all cumulant moments
of the mutual information with order $n>2$ are negligible in the
limit $N\rightarrow \infty$.

As far as the ZF receiver is concerned, recall that (see Section \ref{sec:cum_moments12_ZF})
we can obtain the SINR of the virtual channels of the ZF receiver by setting the parameter $\alpha$ inside
the corresponding SINRs expression of the MMSE receiver to $\infty$.
Specifically, expressing the relation in terms of the matrices
$\Bm_k$, where we have explicitly denoted the dependence on $\alpha$, we have
\begin{eqnarray}\label{eq:app_gammaZF_MMSE_derivation}
\gamma_k^{\zf} &=& \alpha \lim_{\alpha_0\rightarrow\infty}
\hv_k^\herm \Bm_k(\alpha_0) \hv_k
\\ \nonumber
&=& \alpha \lim_{\alpha_0\rightarrow\infty} \hv_k^\herm
\left[\Id+\alpha_0\Hm_k\Hm_k^\herm\right]^{-1} \hv_k
\end{eqnarray}
As mentioned in Section~\ref{sec:cum_moments12_ZF}, the above
analysis involving the matrices $\Bm_{\{k_i\}}$ etc. can be carried
out in the case of zero-forcing if $\beta < 1$. In addition, as seen
in Appendix \ref{app:eig_flucts}, for $\beta<1$ all $n$-th order
cumulant moments of traces of products of $\Bm_{\{k_i\}}$ are given
by (\ref{eq:beenakker_gen_result2}) and
(\ref{eq:beenakker_gen_result3}). As a result, all finite $n>2$
order cumulant moments of the mutual information are $o(1)$ for the
ZF receiver too.

\newpage



\begin{thebibliography}{99}

\bibitem{ieee802.11n}
Draft Standardization Document: IEEE P802.11n/D2.00, February
2007.

\bibitem{Sha}
Claude E.~Shannon, ``Probability of Error for Optimal Codes in a
Gaussian Channel,'' \emph{The Bell System Technical Journal},
vol.~38, no.~3, pp. 611 - 656, May 1959.

\bibitem{ZheTse} L. Zheng and D. Tse, ``Diversity and
multiplexing: A fundamental tradeoff in multiple-antenna
channels,'' {\em IEEE Trans. Info. Theory}, vol. 49, no. 5, pp.
1073-1096, May 2003.

\bibitem{Tel}
I. E.~Telatar, ``Capacity of multi-antenna Gaussian channels,''
\emph{Europ. Trans. Telecomm.}, vol.~10, no.~6, pp. 585--595,
Nov.-Dec. 1999.

\bibitem{vblast}
P. Wolniansky, J. Foschini, G. Golden and R. Valenzuela,
``V-BLAST: an architecture for realizing very high data rates over
the rich-scattering wireless channel,'' Proc. of {\em 1998 URSI
Int. Symp. on Signals, Systems, and Electronics,} pages 295--300,
1998.

\bibitem{wstc} G. Caire and G. Colavolpe, ``On low-complexity space-time coding for quasi-static channels,''
{\em IEEE Trans. on Inform. Theory,} Vol. 49, No. 6, pp.
1400-1416, June 2003.

\bibitem{hesham-iterative}
H. El Gamal and R. Hammons, ``A new approach to layered space-time
coding and signal processing,'' {\em IEEE Trans. on Inform.
Theory,} Vol. 47, No. 6, pp. 2321--2334, June 2001.

\bibitem{VerHan}
S. Verd$\acute{\text{u}}$ and T. S. Han, ``A General Formula for
Channel Capacity,'' \emph{IEEE Trans. on Information Theory}, vol.
40, no. 4, pp. 1147-1157, July 1994.

\bibitem{bps}
E. Biglieri, J. Proakis and S. Shamai, ``Fading Channels:
Information-Theoretic and Communications Aspects,'' {\em IEEE
Trans. on Inform. Theory,} Vol. 44, No. 6, pp. 2619--2692, Oct.
1998.

\bibitem{DamGamCai}
M. O. Damen, H. El Gamal and G. Caire, ``On maximum likelihood
detection and the search for the closest lattice point,'' {\em
IEEE Trans. on Inform. Theory,} Vol. 49, No. 10, pp. 2389-2402,
Oct. 2003.

\bibitem{MurGamDamCai}
A. D. Murugan, H. El Gamal, M. O. Damen, and G. Caire, ``A Unified
Framework for Tree Search Decoding: Rediscovering the Sequential
Decoder,'' {\em IEEE Trans. on Inform. Theory,} Vol.~52, No.~3,
pp. 933 - 953, Mar. 2006.


\bibitem{Ver} Sergio Verd\'{u}, \emph{Multiuser detection},
Cambridge Univ. Press, 1998.

\bibitem{book1}
E. Biglieri and G. Taricco, ``Transmission And Reception With
Multiple Antennas: Theoretical Foundations,'' {\em Foundations and
Trends in Communications and Information Theory}, Now Publishers
Inc., 2004.

\bibitem{book2}
A. Paulraj, R. Nabar, and D. Gore, {\em Introduction to Space-Time
Wireless Communications,} Cambridge University Press, Cambridge
UK, 2003.

\bibitem{TseVis}
D.~Tse and P.~Viswanath, {\em Fundamentals of Wireless
Communication,} Cambridge University Press, Cambridge UK, 2005.

\bibitem{GamCaiDam_LaST}H. El Gamal, G. Caire and M.O.
Damen,``Lattice Coding and Decoding Achieve the Optimal
Diversity-Multilpexing Tradeoff of MIMO Channels,''\textit{IEEE
Trans. Inform. Theory}, Vol.~50, No.~6, pp. 968-985, June 2004.



\bibitem{JiaVar}
Y. Jiang, and M. K. Varanasi, ``Spatial Multiplexing Architectures
with Jointly Designed Rate-Tailoring and Ordered BLAST Decoding -
Part I: Diversity-Multiplexing Tradeoff Analysis,'' {\em IEEE Trans.
Wireless Communication}, Vol.~7, No.~8, pp. 3252 - 3261, Aug. 2008.

\bibitem{JiaVarLi}
Y. Jiang, M. K. Varanasi, and J. Li, ``Performance analysis of ZF
and MMSE equalizers for MIMO systems: A closer study in high SNR
regime,'' {\em IEEE Trans. Inform. Theory}, submitted Aug. 2006, and
accepted for publication.

\bibitem{Nos-isit07}
A. Tajer, A. Nosratinia, and N. Al-Dhahir, ``MMSE Infinite Length
Symbol-by-Symbol Linear Equalization Achieves Full Diversity,'' in
Proc. {\em IEEE International Symposium on Information Theory
(ISIT 2007),} Nice, France, June 24-29, 2007.

\bibitem{HedNos} A.~Hedayat and A.~Nosratinia, ``Outage and Diversity
of Linear Receivers in Flat-Fading MIMO Channels,'' {\em IEEE
Trans. Signal Proc.}, Vol.~55, No.~12, pp. 5868 - 5873, Dec. 2007.

\bibitem{WinSalGit}
J. H. Winters, J.~Salz  and R.~D.~Gitlin, ``The impact of antenna
diversity on the capacity of wireless communication systems,'' {\em
IEEE Trans. Communications}, Vol.~42, No.~234, pp. 1740 - 1751,
1994.

\bibitem{itw07}
K. R. Kumar, G.  Caire and A. L. Moustakas,
``The Diversity-Multiplexing Tradeoff of Linear MIMO Receivers,''
{\em IEEE Information Theory Workshop, ITW' 07,}
pp. 487--492,  2-6 Sept. 2007.

\bibitem{Hochwald2002_MultiAntennaChannelHardening}
B.~M. Hochwald, T.~L. Marzetta, and V.~Tarokh, ``Multi-antenna
channel
  hardening and its implications for rate feedback and scheduling,''
  \emph{{IEEE} Trans. Inform. Theory}, vol.~50, no.~9, pp. 1893--1909, Sept.
  2004.

\bibitem{Moustakas2003_MIMO1}
A.~L. Moustakas, S.~H. Simon, and A.~M. Sengupta, ``\mbox{MIMO}
capacity
  through correlated channels in the presence of correlated interferers and
  noise: \mbox{A} (not so) large \mbox{N} analysis,'' \emph{{IEEE} Trans.
  Inform. Theory}, vol.~49, no.~10, pp. 2545--2561, Oct. 2003.

\bibitem{Smith2002_OnTheGaussianApproximationToTheCapacityOfWirelessMIMOSystem%
s} P.~J. Smith and M.~Shafi, ``On the {G}aussian approximation to
the capacity of
  wireless {MIMO} systems,'' \emph{Proceedings, IEEE International Conference
  on Communications}, p. 406, 2002.

\bibitem{Hachem2007_NewApproachGaussianMIMO}
W. Hachem, O.~Khorunzhiy, P.~Loubaton, J.~Najim and L.~Pastur, ``A new
approach
for mutual information analysis of large dimensional multi-antenna
channels",'' \emph{{IEEE} Trans. Inform. Theory}, vol.~54, no.~9, pp. 3987--2561, Sept. 2008.


\bibitem{Tse2000_MMSEFluctuations}
D.~N. Tse and O.~Zeitouni, ``Linear multiuser receivers in random
  environments,'' \emph{{IEEE} Trans. Inform. Theory}, vol.~46, no.~1, p. 171,
  Jan. 2000.

\bibitem{Liang2007_MMSEAsymptotics}
Y.~C. Liang, G. Pan and Z.~D. Bai, ``Asymptotic Performance of
MMSE Receivers for Large Systems Using Random Matrix Theory,''
\emph{{IEEE} Trans. Inform. Theory}, vol.~53, no.~11, p. 4173,
  Nov. 2007.

\bibitem{Debbah2003_UnitaryAsymptoticallyFreeMatrices}
M.~Debbah \emph{et~al.}, ``{MMSE} analysis of certain large
isometric random
  precoded systems,'' \emph{{IEEE} Trans. Inform. Theory}, vol.~49, no.~5, p.
  1293, May 2003.


\bibitem{KumCai_slast}
K. Raj Kumar and G. Caire, ``Space-Time Codes from Structured
Lattices,'' Accepted for publication in {\em IEEE Trans. Inform.
Theory,} 2008.

\bibitem{viterbo-tcm}
D. Champion, J.-C. Belfiore, G. Rekaya and E. Viterbo,
``Partitionning the Golden Code: A framework to the design of
Space-Time coded modulation,'' {\em Canadian Workshop on Inform.
Theory,} Montreal, 2005.

\bibitem{Men} J. M. Mendel, ``Tutorial on Higher-Order Statistics (Spectra) in
Signal Processing and Systems Theory: Theoretical Results and Some
Applications,'' {\em Proceedings of the IEEE}, vol.~79, No.~3, pp.
278-305, 1991.

\bibitem{Tse1999_LMUReceivers}
D.~N. Tse and S.~Hanly,, ``Linear multiuser receivers: Effective
interference, effective bandwidth and user capacity,''
\emph{{IEEE} Trans. Inform. Theory}, vol.~45, pp. 641–657,
  Mar. 1999.


\bibitem{verdu-shamai99}
S. Verdu and S. Shamai,
``Spectral efficiency of CDMA with random spreading,''
{\em IEEE Trans. on Inform. Theory,}
Vol. 45,  No. 2,  pp.  622 - 640, March 1999.

\bibitem{verdu-shamai01}
S. Verdu and S. Shamai,
``The Impact of Frequency-Flat Fading on the Spectral Efficiency of CDMA,''
{\em IEEE Trans. on Inform. Theory,}
Vol. 47, No. 4, pp. 1302--1327, May 2001



\bibitem{Novikov1965_NovikovTheorem}
Novikov, \emph{Sov. Phys. JETP}, vol.~20, p. 1290, 1965.

\bibitem{Tulino2004_RMTInfoTheoryReview}
A.~M. Tulino and S.~Verd{\'u}, ``Random matrix theory and wireless
  communications,'' \emph{Foundations and Trends in Communications and
  Information Theory}, vol.~1, no.~1, pp. 1--182, 2004.

\bibitem{Beenakker1997_MesoscopicReview}
C.~W.~J. Beenakker, ``Random-matrix theory of quantum transport,''
\emph{Rev.
  Mod. Phys.}, vol.~69, pp. 731--808, 1997.

\bibitem{Politzer1989_RMTCumulants}
D.~H. Politzer, ``Random matrix description of the distribution of
mesoscopic
  conductance,'' \emph{Phys. Rev. {\bf B}}, vol.~40, no.~17, p. 11917, 1989.

\end{thebibliography}
\end{document}